%% file: 1.4.tex
\documentclass{vldb}
\usepackage{graphicx}
\usepackage{balance}  
\usepackage{multirow}
\usepackage{mathrsfs}
\usepackage{amstext}
\usepackage{pbox}
\usepackage{graphicx}
\usepackage{subfigure}
\usepackage{enumerate}
\usepackage{color}
\usepackage{times}
\usepackage[linesnumbered,ruled,vlined]{algorithm2e}

\newcommand{\stitle}[1]{\vspace{1ex} \noindent{\bf #1}}
\newcommand{\eop}{\hspace*{\fill}\mbox{$\Box$}\vspace{0.1cm}}
\newtheorem{definition}{Definition}
\newtheorem{corollary}{Corollary}
\newtheorem{claim}{Claim}

\newtheorem{problem}{Problem}
\newtheorem{theorem}{Theorem}
\newcommand\numcircledmod[1]{\raisebox{.5pt}{\textcircled{\raisebox{-.9pt} {#1}}}}
\newcommand{\proofsketch}{\vspace{-0.2cm}\noindent{\bf Proof Sketch: }}
\newcommand{\ctrack}{{\sffamily cTrack}\xspace}
\newcommand{\etrack}{{\sffamily eTrack}\xspace}
\newcommand{\lsearch}{{\sffamily Linkage Search}\xspace}

\newcommand{\hashtags}{{\sffamily HashtagPeaks}\xspace}
\newcommand{\unigrams}{{\sffamily UnigramPeaks}\xspace}
\newcommand{\louvain}{{\sffamily Louvain}\xspace}
\newcommand{\eventevol}{{\sffamily event evolution tracking}\xspace}

\newcommand{\from}[2]{} 
\newcommand{\comment}[1]{}
 
\newcommand{\squishlist}{ 
 \begin{list}{$\bullet$} 
  {  \setlength{\itemsep}{0pt}
     \setlength{\parsep}{3pt}
     \setlength{\topsep}{3pt}
     \setlength{\partopsep}{0pt}
     \setlength{\leftmargin}{2em}
     \setlength{\labelwidth}{1.5em}
     \setlength{\labelsep}{0.5em}
} }
\newcommand{\squishlisttight}{
 \begin{list}{$\bullet$}
  { \setlength{\itemsep}{0pt}
    \setlength{\parsep}{0pt}
    \setlength{\topsep}{0pt}
    \setlength{\partopsep}{0pt}
    \setlength{\leftmargin}{2em}
    \setlength{\labelwidth}{1.5em}
    \setlength{\labelsep}{0.5em} 
} }

\newcommand{\squishdesc}{
 \begin{list}{}
  {  \setlength{\itemsep}{0pt}
     \setlength{\parsep}{3pt}
     \setlength{\topsep}{3pt}
     \setlength{\partopsep}{0pt}
     \setlength{\leftmargin}{1em}
     \setlength{\labelwidth}{1.5em}
     \setlength{\labelsep}{0.5em}
} }

\newcommand{\squishend}{
  \end{list}
}

\begin{document}

\title{Event Evolution Tracking from Streaming Social Posts}

\numberofauthors{3}
\author{
\alignauthor
Pei Lee\\
       \affaddr{University of British Columbia}\\
       \affaddr{Vancouver, BC, Canada}\\
       \email{peil@cs.ubc.ca}
\alignauthor
Laks V.S. Lakshmanan\\
       \affaddr{University of British Columbia}\\
       \affaddr{Vancouver, BC, Canada}\\
       \email{laks@cs.ubc.ca}
\alignauthor
Evangelos Milios\\
       \affaddr{Dalhousie University}\\
       \affaddr{Halifax, NS, Canada}\\
       \email{eem@cs.dal.ca}
}

\maketitle

\begin{abstract}
Online social post streams such as Twitter timelines and forum discussions have emerged as important
channels for information dissemination. They are noisy, informal, and surge quickly. Real life
events, which may happen and evolve every minute, are perceived and circulated in post streams by
social users. Intuitively, an event can be viewed as a dense cluster of posts with a life cycle
sharing the same descriptive words. There are many previous works on event detection from social
streams. However, there has been surprisingly little work on tracking the evolution patterns of
events, e.g., birth/death, growth/decay, merge/split, which we address in this paper. To define a
tracking scope, we use a sliding time window, where old posts disappear and new posts appear at 
each moment. Following that, we model a social post stream as an evolving network, where each 
social post is a node, and edges between posts are constructed when the post similarity is above a
threshold. We propose a framework which summarizes the information in the stream
within the current time window as a ``sketch graph'' composed of ``core'' posts.  We develop
incremental update algorithms to handle highly dynamic social streams and track event evolution
patterns in real time. Moreover, we visualize events as word clouds to aid human perception. Our
evaluation on a real data set consisting of 5.2 million posts demonstrates that our method can
effectively track event dynamics in the whole life cycle from very large volumes of social streams
on the fly.
\end{abstract}

\section{Introduction}\label{intro}

In the current social web age, people easily feel overwhelmed by the 
information deluge coming from post streams which flow in from channels like
Twitter, Facebook, forums, Blog websites and email-lists.
As of 2009, it was reported \cite{twitter-user-report}, e.g.,
that each Twitter user follows 126 users on average, and on each day, the received social
streaming posts will cost users considerable time to read, only to discover a small
interesting part. This is a huge overhead that users pay in order to find a small amount
of interesting information. There is thus an urgent need to provide users with tools which
can automatically extract and summarize significant information from highly dynamic social streams,
e.g., report emerging bursty events, or track the evolution
of a specific event in a given time span.
There are many previous studies
\cite{kdd09-LeskovecBK,chi11-MarcusBBKMM,www10-SakakiOM,wsdm11-SarmaJY,vldb05-FungYYL,
wsdm10-BeckerNG} on detecting new emerging events from text streams; they serve the need for answering the query
``\emph{what's happening?}''  over social streams.  However, in many scenarios, users may
want to know more details about an event and may like to issue advanced queries like
``\emph{how're things going?}''. For example, for the event ``SOPA (Stop Online Piracy Act)
protest'' happening in January 2012, existing event detection approaches can discover 
bursty activities at each moment, but cannot answer queries like ``how SOPA protest has
evolved in the past few days?''. An ideal output to such an evolution query would be a
``panoramic view'' of the event history, which improves user experience. In this work, we consider
this kind of queries as an instance of the \eventevol problem, which aims to track the
event evolutionary dynamics at each moment from social streams. Typical event
evolution patterns include emerging (birth) or disappearing (death), inflating (growth) or shrinking
(decay), and merging or splitting of events. Event detection can be viewed as a subproblem of event
evolution tracking. We try to solve the \eventevol problem in the paper.

There are several major challenges in the tracking of event evolution.  The first
challenge is the effective organization of noisy social post streams into a meaningful
structure. Social posts such as tweets are usually written in an informal way, with lots
of abbreviations, misspellings and grammatical errors. Even worse, a correctly written
post may have no significance and be just noise. Recent works on event detection from
Twitter \cite{icwsm11-WengL,chi11-MarcusBBKMM} recognize and handle noise in post streams
in a limited and ad hoc manner but do not handle noise in a systematic formal
framework. The second challenge is how to track and express the event evolution behaviors
precisely and incrementally. Most related work reports event activity by volume on the
time dimension \cite{chi11-MarcusBBKMM,kdd09-LeskovecBK}. While certainly useful, this
cannot show the evolution behaviors about how events are split or merged, for instance.
The third challenge is the summary and annotation of events. Since an event may easily
contain thousands of posts, it is important to summarize and annotate it in order to
facilitate human perception. Recent related works
\cite{chi11-MarcusBBKMM,kdd09-LeskovecBK,icwsm11-WengL} typically show users just a list
of posts ranked by importance or time freshness, which falls short of addressing this
challenge.

\begin{figure*}[t]
\centering
\subfigure[The sketch-based framework]{
  \includegraphics[width=8cm] {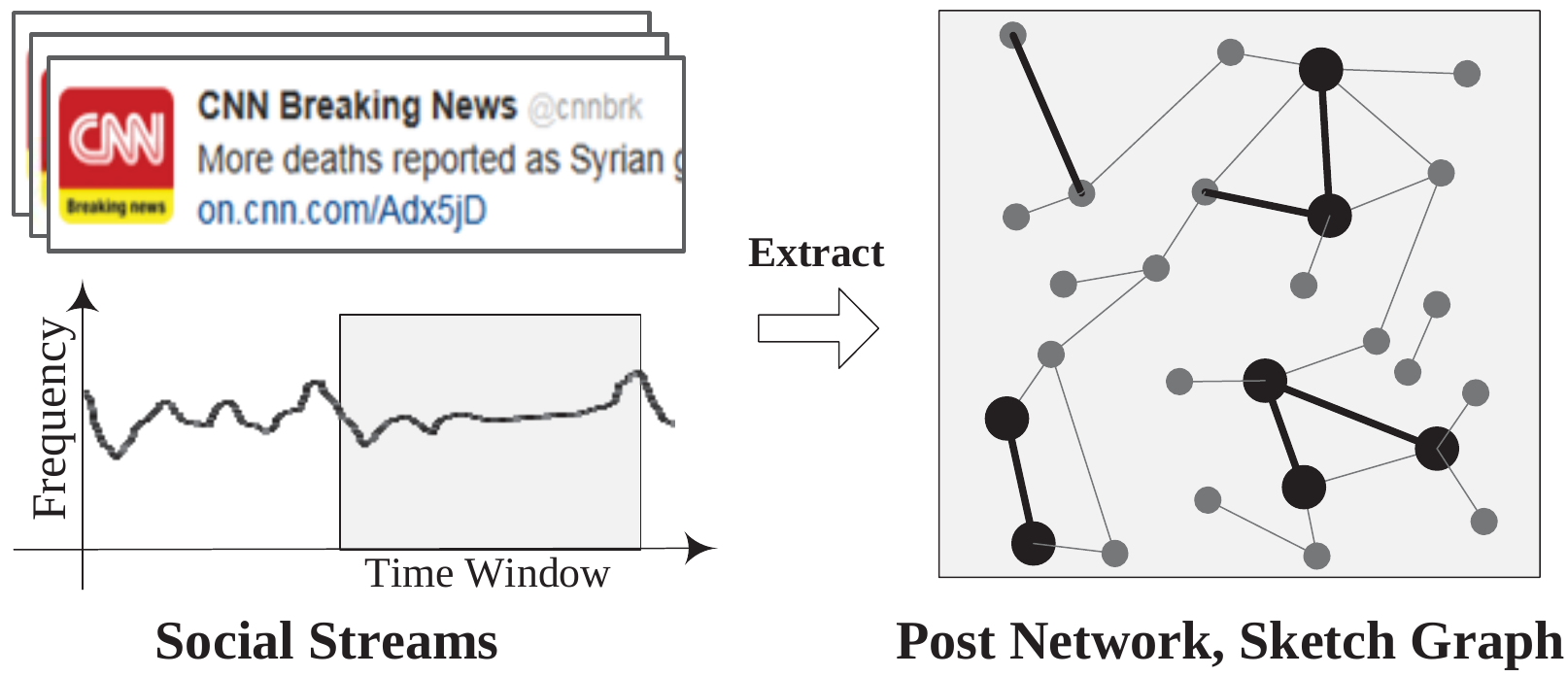}
  \label{overview:a}
}\hspace{1cm}
\subfigure[Event evolution tracking]{
  \includegraphics[width=8cm] {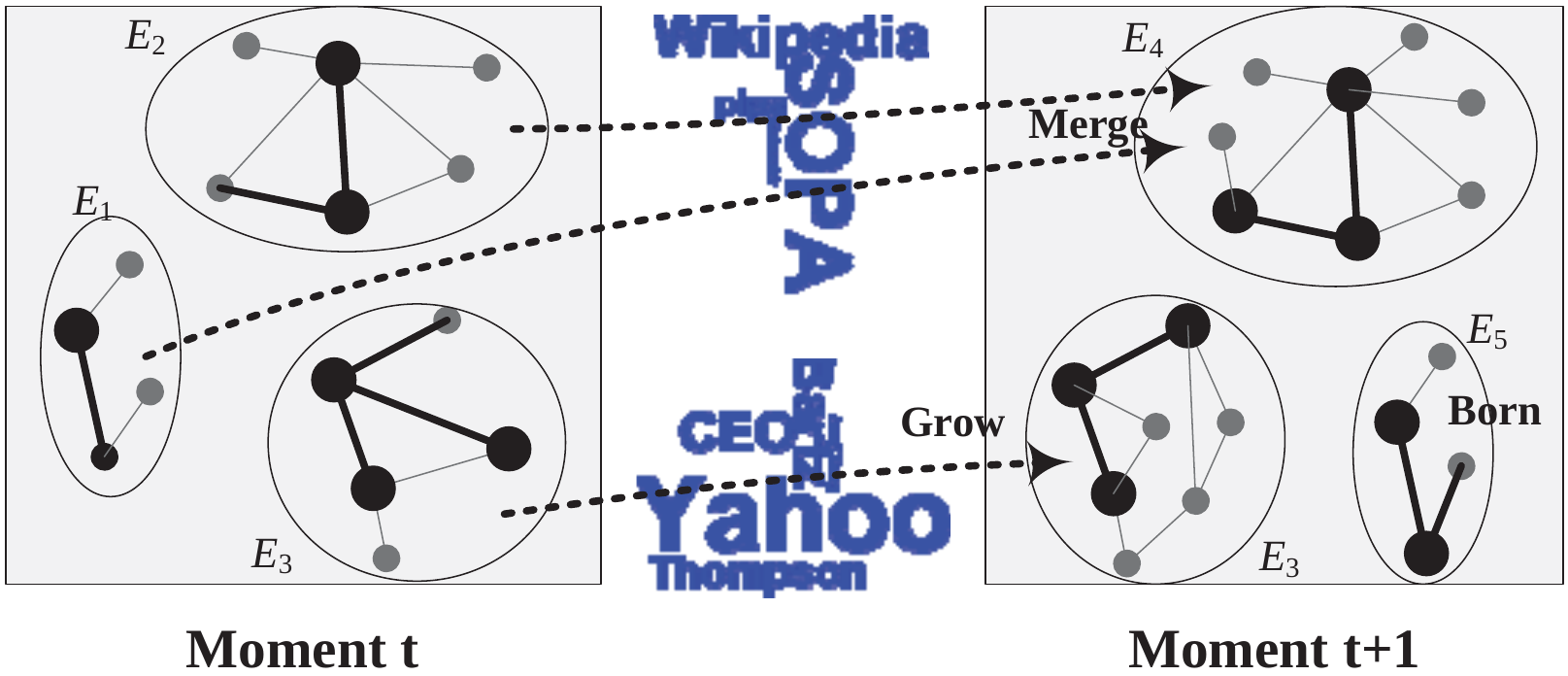}
  \label{overview:b}
}
\vspace{-0.3cm}
\caption{(a) Post network captures the correlation between posts in
the time window at each moment, and evolves as time rolls on. Core
nodes and core edges are bolded. (b) From moment $t$ to $t+1$, typical
event evolution behaviors include birth or death, growth
or decay, merging or splitting of events. Each event is annotated by a
topical word cloud.}\label{overview}
\vspace{-0.2cm}\end{figure*}

{To handle the above mentioned challenges,} we first model social streams as an
\emph{evolving post network}, and then propose a \emph{sketch graph}-based framework to track
the evolution of events in the post network (Section~\ref{sec:sketch}). Intuitively, a
sketch graph can be viewed as a compact summary of the original post network. The
sketch graph only contains \emph{core} posts from the post network, where core posts are
defined as posts that play a {\em central role} in the network. Noise posts will be directly
pruned by the sketch graph. As we will
discuss in Section~\ref{s:tracking}, evolution behaviors can be effectively and incrementally expressed based on a group of primitive operations in the
sketch graph. Technically, we define an event as a cluster in the post network, and then summarize
and annotate each event by topical word-clouds.  
We show an overview of major steps for event tracking from social streams in Figure
\ref{overview}. Note that as time rolls on, the post network, events and their annotations
will be updated incrementally at each moment.

\comment{Here, the term sketch means core
posts that play an central role, and the induced sketch graph can be
viewed as a summary or compression of the original post network.}

We notice that at a high level, our definition seemingly resembles
previous work on density-based clustering over streaming data, e.g., DenStream in \cite{sdm06-CaoEQZ}
 and cluster maintenance in \cite{pvldb12-AgarwalRB} and \cite{pvldb12-AngelKSS}.
However, there are several major differences. First, our approach works on an evolving graph structure and provides
users the flexibility in choosing the scope for tracking and monitoring new events by
means of a fading time window, while the existing work doesn't
provide a tracking scope. Second, the existing work can only process the adding of nodes/edges one
by one, while our approach can handle adding, deleting and fading of nodes {\sl subgraph by subgraph}. This
is an important requirement for dealing with the high throughput rate of online post
streams. Third, the focus of our approach is tracking and analyzing the evolution
dynamics in the whole life cycle of events. By contrast, they focus on the
maintenance of clusters, which is only a sub-task in our problem.

Finally, to compare with traditional topic tracking on news article streams, we note that
this problem is usually formulated as a classification problem \cite{Allan-2002-TDT}: when
a new story arrives, compare it with topic features in the training set and if it matches
sufficiently, declare it to be on a topic. Commonly used techniques include decision trees
and $k$-NN \cite{sigir00-YangAPL}. This approach assumes that topics are {\sl predefined
  before tracking}.  Thus, we cannot simply apply topic tracking techniques to event
tracking in social streams, since future events are unknown and may not conform to any
previously known topics.  Moreover, traditional topic tracking has difficulties in
tracking the composite behaviors such as merging and splitting, which are definitely a key
aspect of event evolution.

In summary, the main problem we study in this paper is captured by the following
questions: how to efficiently track the evolution behavior of social post streams such as
Twitter, which are noisy and highly dynamic? how to do this incrementally?  what is an
effective way to express evolution behavior of events?  In this paper, we develop a
framework and algorithms to answer all these questions. Our main contributions are the
following:

\squishlisttight 
\item We design an effective approach to extract and organize
meaningful information from noisy social streams (Section \ref{pnc});

\item We filter noisy posts by introducing
\emph{sketch graph} (Section \ref{sec:sketch})), based on which we define a group of
primitive operations and express evolution behaviors with respect to graphs and
events using these primitive operations (Section \ref{s:tracking});

\item We propose efficient algorithms \ctrack and \etrack to track the evolution of
clusters and events accurately and incrementally, superior in both quality
and performance to existing approaches that find the evolution patterns by matching events
in consecutive time moments (Section \ref{s:alg}).

\item Our evaluation on a large real data set demonstrates 
that our method can effectively track all six kinds of event evolution behaviors from
 highly dynamic social post streams in a scalable manner (Section \ref{s:exp}).
\squishend 

More related work is discussed in Section~\ref{sec:related}. We summarize the paper and discuss
extensions in Section~\ref{s:con}. For convenience, we summarize the major notations used in the paper, in Table \ref{notations}.

\begin{table}[t]\label{notations}
\centering
{\small\noindent
\begin{tabular}{|c|c|}
\hline
$S(p_1,p_2)$ & the content similarity between $p_1$ and $p_2$\\\hline
$S_F(p_1,p_2)$ & the fading similarity between $p_1$ and $p_2$\\\hline
$(\delta_1,\varepsilon_0,\varepsilon_1)$ & density factors for core
post, edge and core edge\\\hline
$w^{t}(p)$ & the weight of post $p$ at time moment $t$\\\hline
$\mathcal{G}_t(\mathcal{V}_t,\mathcal{E}_t)$ & the post network at
moment $t$\\\hline
$G_t(V_t,E_t)$ & the sketch graph of the post network at moment $t$\\\hline
$\mathcal{N}(p)$ & the neighbor set of $p$ in $\mathcal{G}_t$\\\hline
$\overline{C}, \overline{S}_t$ & a component, a component set at
moment $t$\\\hline
$C,S_t$ & a cluster, a cluster set at moment $t$\\\hline
$N_c(p)$ & the set of clusters where $p$'s neighboring core posts
belong\\\hline
$E, \widetilde{S}_t$ & an event, an event set at moment $t$\\\hline
$A$ & the annotation for an event\\\hline
\end{tabular}
}
\vspace{-0.2cm}\caption{Notation.}
 \vspace{-0.3cm}
\end{table}

\input{relatedwork}

\input{postnetwork}

\input{sketch}

\input{tracking}

\input{alg}

\input{exp}

\section{Conclusion}\label{s:con}
Our main goal in this paper is to track the evolution of events over social streams such
as Twitter. To that end, we extract meaningful information from noisy post streams
and organize it into an evolving network of posts under a sliding time window. We model
events as sufficiently large clusters of posts sharing the same topics,
and propose a framework to describe event evolution behaviors using a set of primitive operations. Unlike previous approaches, our evolution tracking algorithm
\etrack performs incremental updates and efficiently tracks event evolution patterns in
real time. We experimentally demonstrate the performance and quality of our algorithm over
two real data sets crawled from Twitter.
As a natural progression, in the future, it would be interesting to investigate the tracking of
evolution of social emotions on products, with its obvious application for business intelligence.

\bibliographystyle{abbrv}
\bibliography{vldb}

\end{document}

%% file: relatedwork.tex
\section{Related Work}\label{sec:related}

Work related to this paper mainly falls in one of these categories.

\stitle{Topic/Event/Community detection and tracking}.
Most previous works detect events by discovering topic bursts from a document stream. 
Their major techniques are either detecting the frequency peaks of event-indicating phrases over
time in a histogram, or monitoring the formation of a cluster from a structure perspective.
A feature-pivot clustering is proposed in \cite{vldb05-FungYYL} to detect bursty events
from text streams. Sarma et al. \cite{wsdm11-SarmaJY} design efficient algorithms to
discover events from a large graph of dynamic relationships. Jin et al. \cite{www10-JinSMH} present Topic Initiator Detection (TID) to
automatically find which web document initiated the topic on the Web. Louvain method
\cite{louvain}, based on modularity optimization, is the state-of-the-art community
detection approach which outperforms others. However, Louvain method cannot not resist massive
noise. None of the above works address the event evolution tracking problem.

There is less work on evolution tracking.
An event-based characterization of behavioral patterns for communities in temporal
interaction graphs is presented in \cite{kdd07-AsurPU}. A framework for
tracking short, distinctive phrases (called ``memes'') that travel relatively intact
through on-line text was developed in \cite{kdd09-LeskovecBK}. 
Unlike these works, we focus on the tracking of real world event-specific evolution patterns from
social streams.

\stitle{Social Stream Mining}. 
Weng et al. \cite{icwsm11-WengL} build signals for
individual words and apply wavelet analysis on the frequency of words to detect
events from Twitter.
Twitinfo \cite{chi11-MarcusBBKMM} detects events by keyword peaks and 
 represents an event it discovers from Twitter by a timeline of related tweets. Recently, Agarwal et
 al. \cite{pvldb12-AgarwalRB} discover events that are unraveling in microblog streams, by modeling
 events as dense clusters in highly dynamic graphs. Angel et al. \cite{pvldb12-AngelKSS} study the
 efficient maintenance of dense subgraphs under streaming edge weight updates.
Both \cite{pvldb12-AgarwalRB} and \cite{pvldb12-AngelKSS} model the social stream as an evolving
entity graph, but suffer from the drawback that post attributes like time and author cannot be reflected. Another
drawback of \cite{pvldb12-AgarwalRB} and \cite{pvldb12-AngelKSS} is that they can only handle 
edge-by-edge updates, but cannot handle subgraph-by-subgraph bulk updates. Both drawbacks are solved
in our paper.

\stitle{Clustering and Evolving Graphs}.
In this paper, we summarize an original post network into a sketch graph based on density
parameters. Compared with partitioning-based
approaches (e.g., K-Means \cite{dmbook}) and
hierarchical approaches (e.g., BIRCH \cite{dmbook}), density-based clustering (e.g., DBSCAN \cite{dmbook}) is effective in finding arbitrary-shaped
clusters, and is robust to noise. The main challenge is to apply density-based clustering on
fast evolving post networks. CluStream \cite{vldb03-AggarwalHWY} is a framework
that divides the clustering process into an online component which periodically generates
detailed summary statistics for nodes and an offline component which uses only the summary
statistics for clustering. However, CluStream is based on K-Means only.
{DenStream \cite{sdm06-CaoEQZ} presents a new approach
for discovering clusters in an evolving data stream by extending DBSCAN.}
This work is related to us in that both employ
density based clustering. The differences between our approach and DenStream were discussed in
detail in the introduction. Subsequently,
DStream \cite{kdd07-ChenT} uses an online component which maps each input data record into a
grid and an offline component which generates grid clusters based on the density.
Another related work is by Kim et al. \cite{pvldb09-KimH}, which first clusters
individual snapshots into quasi-cliques and then maps them over time by looking at the density of
bipartite graphs between quasi-cliques in adjacent snapshots. Although \cite{pvldb09-KimH} can
handle birth, growth, decay and death of clusters, the splitting and merging behaviors are not
supported. In contrast, our approach is totally incremental and is able to track composite
behaviors like merging and splitting in real time.

%% file: postnetwork.tex
\section{Post Network Construction}\label{pnc}

In this section, we describe how we construct a post network
from a social post stream. The main challenge is detecting similarity between 
streaming posts
efficiently and accurately, taking the time of the posts into account.
We use a notion of \emph{fading similarity} (Section~\ref{s:fs})
and propose a technique called \emph{linkage search} to efficiently detect posts
similar to a post as it streams in (Section~\ref{s:network}).

\comment{
In this section, we first introduce the preprocessing techniques of
social streams and the definition of fading similarity between posts.
Then we discuss the construction of post network by linkage search
while the posts are streaming in.
}

\subsection{Social Stream Preprocessing}
\label{s:ssp}

Social posts such as tweets are usually written in an informal way.
Our aim is to design a processing strategy that can quickly judge what
a post talks about and is robust enough to the informal writing style.
In particular, we focus on the entity words contained in a post, since
entities depict the topic. For example, given a tweet ``iPad 3 battery
pointing to thinner, lighter tablet?'', the entities are ``iPad'',
``battery'' and ``tablet''. However, traditional Named Entity
Recognition tools \cite{LI07-Nadeau} only support a
narrow range of entities like Locations, Persons and Organizations.
NLP parser based approaches
\cite{acl03-KleinM} are not appropriate due to the informal writing style
of posts and the need for high processing speed. Also, simply treating
each post text as a bag of words \cite{IRBook} will lead to loss of
accuracy, since different words have different weights in deciding the
topic of a post. To broaden the applicability, we treat each noun in
the post text as a candidate entity. Technically, we obtain nouns
from a post text using a Part-Of-Speech Tagger\footnote{POS Tagger,
http://nlp.stanford.edu/software/tagger.shtml}, and if a noun is
plural (POS tag ``NNS'' or ``NNPS''), we obtain the prototype of this
noun using WordNet stemmer\footnote{JWI WordNet Stemmer,
http://projects.csail.mit.edu/jwi/}. In practice, we find this
preprocessing technique to be robust and efficient. In the Twitter
dataset we used in our experiments (see Section~\ref{s:exp}), each tweet
contains 4.9 entities on an average.
We formally define a social post as follows.

\begin{definition}\label{post} \textup{({Post})}. A post $p$ is
a triple $(L,\tau,u)$, where $L$ is the list of entities in the post,
$\tau$ is the time stamp of the post, and $u$ is the user who posted it.
\end{definition}\vspace{-0.2cm}

We let $p^L$ denote $L$ in the post $p$ for simplicity, and analogously for
$p^\tau$ and $p^u$. We use $|p^L|$ to denote the
number of entities in $p$.

\subsection{Fading Similarity}
\label{s:fs}

Post similarity is the most crucial criterion in correlating posts of
the same event together. Traditional similarity measures such as
TF-IDF based cosine similarity, Jaccard Coefficient and Pearson
Correlation \cite{IRBook} only consider the post content. However,
clearly time dimension should play an important role in determining post
similarity, since posts created closer together in time are more
likely to discuss the same event than posts created at very different
moments. We introduce the notion of fading similarity to capture both
content similarity and time proximity. Formally, we define the \emph{fading
similarity} between a pair of posts $p_i$ and $p_j$ as
\begin{small}
\vspace{-0.2cm}
\begin{equation}\label{fadingsim}
S_F(p_i,p_j)= {S(p_i^L,p_j^L)\over D(|p_i^\tau-p_j^\tau|)}
\vspace{-0.2cm}
\end{equation}
\end{small}

\noindent where $S(p_i^L,p_j^L)$ is a set-based similarity measure
that maps the similarity between $p_i^L$ and $p_j^L$ to the interval
$[0,1]$, and $D(|p_i^\tau-p_j^\tau|)$ is a distance measure that is
monotonically increasing with $|p_i^\tau-p_j^\tau|$ and
$D(|p_i^\tau-p_j^\tau|)\geq 1$. For example, $S(p_i^L,p_j^L)$ may be the
Jaccard coefficient with $S(p_i^L,p_j^L)={(|p_i^L\cap
p_j^L|)}/{(|p_i^L\cup p_j^L|)}$, and $D(|p_i^\tau-p_j^\tau|)$ may be 
$D(|p_i^\tau-p_j^\tau|)=e^{|p_i^\tau-p_j^\tau|}$. We will compare different
 measures in experiments. It is known that nouns are usually more topic
relevant than verbs, adjectives, etc \cite{www03-LiuCN}. Consequently, entity-based similarity of posts is
more appropriate than similarity based on all words. Besides, it
has the advantage of smaller computational overhead. It is trivial to
see that $0\leq S_F(p_i,p_j)\leq 1$ and that $S_F(p_i,p_j)$ is
symmetric.

\subsection{Post Network and Linkage Search}\label{s:network}

To find the implicit correlation between posts as they stream in,
we build a post network based on the
following rule: if the fading similarity between two posts is higher
than a given threshold $\varepsilon_0$, we create an edge between
them. More formally:

\vspace{-0.2cm}\begin{definition}\label{postnetwork} \textup{({Post Network})}. The
snapshot of \emph{post network} at moment $t$ is defined as a graph
$\mathcal{G}_t(\mathcal{V}_t,\mathcal{E}_t)$, where each node
$p\in\mathcal{V}_t$ is a post, and an edge $(p_i,
p_j)\in\mathcal{E}_t$ is constructed iff
$S_F(p_i,p_j)\geq\varepsilon_0$,
where $0<\varepsilon_0<1$ is a given parameter.
\vspace{-0.2cm}\end{definition}

We monitor social streams using a \emph{fading time window}, which will be
introduced explicitly in Section \ref{timewindowmodel}. For now,
imagine a time window of observation and consider the post network
at the beginning of the time window. As we move forward in time, new posts
flow in and old posts fade out and
$\mathcal{G}_t(\mathcal{V}_t,\mathcal{E}_t)$ is dynamically updated at
each moment, with new nodes/edges added and old nodes/edges removed.
Removing a node and associated edges from
$\mathcal{G}_t(\mathcal{V}_t,\mathcal{E}_t)$ is an easy operation, so let us
investigate the case when a node is added. When a new post $p$ flows in, we need to construct
the linkage (i.e., edges) between $p$ and nodes in $\mathcal{V}_t$, following 
Definition~\ref{postnetwork}.
 In a real scenario, since the number of nodes
$|\mathcal{V}_t|$ can easily go up to millions, it is impractical to
compare $p$ with every node in $\mathcal{V}_t$ to verify the
satisfaction of Definition \ref{postnetwork}. Below, we propose \lsearch to
 identify the neighbors of $p$ accurately by accessing only a small
subset of nodes in $\mathcal{V}_t$.

\stitle{Linkage Search}.
Let $\mathcal{N}(p)$ denote the set of neighbors of $p$ that satisfy Definition
\ref{postnetwork}. The problem of linkage search is finding  $\mathcal{N}(p)$
accurately by accessing only a small node set $\mathcal{N}'$, where
$\mathcal{N}(p)\subseteq \mathcal{N}'\subset \mathcal{V}_t$ and
$|\mathcal{N}'|\ll |\mathcal{V}_t|$. To solve this problem, first we
construct a post-entity bipartite graph, and then perform a two-step
random walk process on this bipartite graph and use the hitting counts. The main idea
of linkage search is to let a random surfer start from post node $p$ and
walk to any entity node in $p^L$ on the first step, and continue
to walk back to post nodes except $p$ on the second step. All the post
nodes visited on the second step form a set $\mathcal{N}'$. For any
node $q\in\mathcal{N}'$, $q$ can be hit multiple times from different
entities and the total hitting count $\mathcal{H}(p,q)$ can be
aggregated. Assuming the measure of $S(p^L,q^L)$ in Eq.~(\ref{fadingsim}) is Jaccard
coefficient, we can verify the linkage between $p$ and $q$ by
checking the condition
\begin{small}\vspace{-0.1cm}
\begin{equation}\label{iceberg}
S_F(p,q)={\mathcal{H}(p,q)\over
{\left(|p^L|+|q^L|-\mathcal{H}(p,q)\right)D(|p^{\tau}-q^{\tau}|)}}\geq
\varepsilon_0
\vspace{-0.2cm}\end{equation}
\end{small}

Linkage search supports the construction of a post network \emph{on the fly}. It is
easy to see that for post $q\not\in\mathcal{N}'$, $S_F(p,q)=0$. Thus,
we do not need to access posts that are not in $\mathcal{N}'$. Since
a post like tweet usually connects to  very few entities,
$|\mathcal{N}'|\ll |\mathcal{V}_t|$ generally holds, thus making linkage
search very efficient. {Suppose the average number of entities in each
post is $d_1$ and the average number of posts mentioning each entity is $d_2$. Then 
linkage search can find the neighbor set of a given post in time $O(d_1d_2)$}.

%% file: sketch.tex
\section{Sketch-Based Summarization}\label{framework}

Here we first introduce the notion of \emph{sketch graph} based on density parameters,
and define events in the post network based on components of the sketch
graph. The relationships between different types of objects defined in
this paper are illustrated in Figure \ref{relationships}. 
{As an example to explain this figure, the arrow from $\mathcal{G}_t$ to $G_t$ labeled
$Ske$ means $G_t=Ske(\mathcal{G}_t)$. See Table \ref{notations} for notations. }

\subsection{Sketch Graphs}\label{sec:sketch}

Many posts tend to be just noise so it is essential to identify those posts that play a central role in describing events. We formally capture them using the notion of \emph{core posts}.
On web link graphs,
there is a lot of research on node authority ranking, e.g., HITS
 and PageRank \cite{IRBook}. However, most of
these methods are iterative and not applicable to one-scan computation
on streaming data.

\begin{figure}[t]
\centering
\includegraphics[width=7cm]{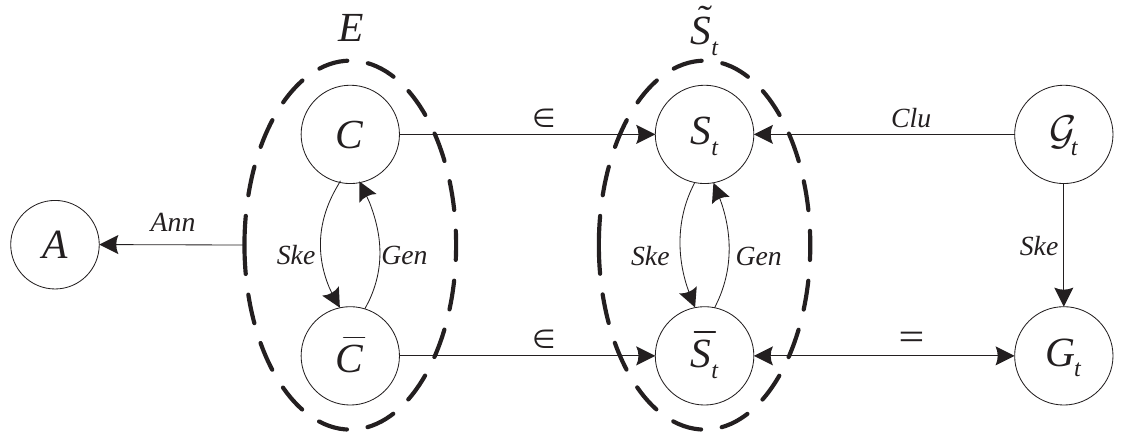}
\vspace{-0.2cm}\caption{The functional relationships between different types of
objects defined in this paper. Refer to Table \ref{notations} for
notations. }\label{relationships}
\vspace{-0.2cm}
\end{figure}

In this paper, we introduce a sketch graph $G_t(V_t,E_t)$ of
 post network $\mathcal{G}_t(\mathcal{V}_t,\mathcal{E}_t)$
based on density parameters $(\delta_1,\varepsilon_0,\varepsilon_1)$,
where $\varepsilon_1\geq\varepsilon_0$. In density-based clustering
, $\delta_1$ (a.k.a. $MinPts$ in DBSCAN
\cite{kdd96-EsterKSX}) is the minimum number of nodes in an
$\varepsilon_0$-neighborhood, required to form a cluster. In the post
network, we consider $\delta_1$ to be the threshold to judge whether a post is
important and similarly $\varepsilon_1$ is a threshold for core edges. The
reason we choose density-based clustering is that, compared with
partitioning-based approaches (e.g., K-Means \cite{dmbook}) and
hierarchical approaches (e.g., BIRCH \cite{dmbook}), density-based
clustering (e.g., DBSCAN) defines clusters as
areas of higher density than the remainder of the data set, which is
effective in finding arbitrary-shaped clusters and is robust to noise.
We adapt the concepts from density-based clustering and define the
post weights as follows.

\begin{definition}\label{d:postweight} \textup{({Post Weight})}. Given a
post $p=(L,\tau,a)$ in $\mathcal{G}_t(\mathcal{V}_t,\mathcal{E}_t)$
and its neighbor set $\mathcal{N}(p)$, the weight of $p$ at moment $t$,
$\;t\geq p^{\tau}$, is defined as
\vspace{-0.2cm}\begin{small}
\begin{equation}\label{core}
w^{t}(p)=\frac{1}{D(|t-p^{\tau}|)}\sum_{q\in \mathcal{N}(p)}{S_F(p,q)}
\vspace{-0.2cm}\end{equation}
\end{small}
\end{definition}

Notice, post weight decays as time moves forward, although
the neighbor set $\mathcal{N}(p)$ may change. Thus, post
weights need to be continuously updated. In practice, we only store the sum
$\sum_{q\in \mathcal{N}(p)}{S_F(p,q)}$ with $p$ to avoid frequent
updates and compute $w^{t}(p)$ on demand when we need to judge the importance
of a post. Based on post weight, we distinguish nodes in
$\mathcal{G}_t(\mathcal{V}_t,\mathcal{E}_t)$ into three types, as defined below.

\vspace{-0.1cm}\begin{definition}\label{d:nodetype} \textup{({Node Types})}.
\squishlisttight 
\item A post $p$ is a \emph{core post} if $w^{t}(p)\geq \delta_1$;
\item It is a \emph{border post} if  $w^{t}(p)<\delta_1$ but there exists at least one
core post $q\in \mathcal{N}(p)$;
\item It is a \emph{noise post} if it is neither core nor border,
i.e., $w^{t}(p)<\delta_1$ and there is no core post in $\mathcal{N}(p)$.
\squishend 
\end{definition}

Intuitively, a post is a core post if it shares enough common entities with
many other posts. Neighbors of a core post are at least border
posts, if not core posts themselves. In density-based clustering, core posts play a central role: if a core post $p$ is found to be a part of a cluster $C$, its
neighboring (border or core) posts will also be a part of $C$.
Analogouly to the notion of core posts, we use the threshold
$\delta_1$ to define \emph{core edges}. An edge
$(p,q)$ is a core edge if $S_F(p,q)\geq\varepsilon_1$, where
$\varepsilon_1\geq\varepsilon_0$. Notice that a core edge may connect
non-core nodes.
\comment{
To make
the clusters more compact, for a neighboring core post $q$, $p$ can
``activate'' $q$ only if they are connected by a core edge. An edge
$(p,q)$ is a core edge if $S_F(p,q)\geq\varepsilon_1$, where
$\varepsilon_1\geq\varepsilon_0$, and $p,q$ are not necessarily core
posts.
}
Core posts connected by core edges will build a summary of
$\mathcal{G}_t(\mathcal{V}_t,\mathcal{E}_t)$, that we call the
sketch graph (See Figure \ref{relationships}).

\begin{definition}\label{d:sketch} \textup{({Sketch Graph})}. Given post network
$\mathcal{G}_t(\mathcal{V}_t,\mathcal{E}_t)$ and density parameters
$(\delta_1,\varepsilon_0,\varepsilon_1)$, we define the \emph{sketch graph} 
of $\mathcal{G}_t$ as the subgraph induced by its core posts and core 
edges. More precisely, the sketch graph $G_t(V_t,E_t)$ satisfies the  
condition that each node
$p\in V_t$ is a core post and each edge $(p,q)\in E_t$ is a core edge. 
We write $G_t=Ske(\mathcal{G}_t)$.
\end{definition}

Intuitively, all important nodes in
$\mathcal{G}_t(\mathcal{V}_t,\mathcal{E}_t)$ and their relationships
are retained in $G_t(V_t,E_t)$. Empirically, we found that adjusting the
granularity of $(\delta_1,\varepsilon_0,\varepsilon_1)$ to make the
size $|V_t|$ roughly equal to 20\% of $|\mathcal{V}_t|$ leads to a good
balance between the quality of the sketch graph in terms of the information
retained and its space complexity. The tuning detail can be found in Section \ref{s:tuning}.

\subsection{Event Identification and Annotation}

We define events based on post clusters.
Recall that 
if two core posts are connected by core edges, they should belong to the same cluster. It implies all core posts in cluster $C$
form a connected component $\overline{C}$ in $G_t(V_t,E_t)$ and we
write $\overline{C}=Ske(C)$. Let $\overline{S_t}$ denote the set of
connected components in $G_t$. Notice that $\overline{S_t}$ and $G_t$ have
the same node set and the same structure, so we 
write $\overline{S_t}=G_t$ for simplicity. To generate clusters based on the sketch graph 
discussed above, we can start from a connected component
$\overline{C}\in\overline{S_t}$ to build a cluster $C$. See Figure \ref{relationships} for
the functional relationships between different types of objects. We define cluster $C$ as follows.

\begin{definition}\label{d:cluster} \textup{({Cluster})}. Given
$\mathcal{G}_t(\mathcal{V}_t,\mathcal{E}_t)$ and the corresponding  sketch graph
$G_t(V_t,E_t)$, a cluster $C$  of $\mathcal{G}_t$ is a set of core posts and border posts
generated from a connected component $\overline{C}$ in $G_t(V_t,E_t)$, written
as $C=Gen(\overline{C})$, and defined as follows:

\squishlisttight 
\item All posts in the component $\overline{C}$ form the core posts of $C$.

\item For every core post in $C$, all its neighboring border posts
in $\mathcal{G}_t$ form the border posts in $C$. 
\squishend 
\vspace{-0.2cm}\end{definition}

A core post only appears in one cluster (by definition). If a border post is
associated with multiple core posts in different clusters, this border
post will appear in multiple clusters. Events are defined based on
clusters in 
$\mathcal{G}_t$. However, not every cluster can form an
event; we treat clusters with a very small size as
outliers, since they do not gain wide
popularity at the current moment. We use $E$ and $O$ to denote an event and an outlier
respectively.

\begin{definition}\label{d:event} \textup{({Event})}. Given a cluster
$C$, the following  function 
distinguishes between an {event} and an {outlier}.
\begin{small}\vspace{-0.1cm}
\begin{equation}\label{e:event}
Event(C)=\left\{ \begin{matrix}
  {\bf true}, \text{ if }|C|\ge \varphi   \\
  {\bf false}, \text{ otherwise}  \\
\end{matrix} \right.
\vspace{-0.2cm}\end{equation}
\end{small}
where $\varphi$ is a size threshold.
\vspace{-0.2cm}\end{definition}

We empirically set $\varphi=10$ for event identification. Note that outliers are different from
noise; noise is associated at the level of posts, as opposed to clusters. Besides, {\sl an outlier at the
current moment may grow into an event in the future, and an event may
also degrade into an outlier as time passes}.

\stitle{Entity Annotation}. Considering the huge volume of posts in an
event, it is important to summarize
and present a post cluster as a conceptual event to aid human perception.
In related work, Twitinfo \cite{chi11-MarcusBBKMM} represents an event
it discovers from Twitter by a timeline of tweets. Although the tweet
activity by volume over time is shown, it is tedious for
users to read tweets one-by-one to figure out the event detail. In
this paper, we summarize a snapshot of an event by a word cloud
\cite{www07-HalveyKa}. The font size of a word in the cloud indicates
its popularity. Compared with Twitinfo, a word cloud provides a
summary of the event at a glance and is much easier for a human to
perceive. Technically, given an event $E$, the annotation $A$ for $E$, computed as $A=Ann(E)$,
is a set of entities with popularity score, expressed as
$A=\{(e_1,P_{e_1}),(e_2,P_{e_2}),\cdots\}$.

We take entities as the word candidates in the cloud. Intuitively, an
entity $e$ is popular in an event if many important posts in this
event contain $e$ in their entity list. Formally, the relationship
between posts and entities can be modeled as a bipartite graph. Recall
that the HITS algorithm \cite{jacm99-Kleinberg} computes the hub score and
authority score by mutual reinforcement: the hub score $\mathcal{H}_i$
of a node $i$ is decided by the sum of authority scores of all nodes
pointed by $i$, and simultaneously, the authority score
$\mathcal{A}_j$ of a node $j$ is decided by the sum of hub scores of
all nodes pointing to $j$. Inspired by HITS, we let the post weight
obtained from the post network be the initial hub score of posts, and
then the authority scores of entities can be computed by one iteration:
$\mathcal{A}=M^T\mathcal{H}$,
where $M$ is an adjacency matrix between posts and entities in an
event, $\mathcal{A}=[P_{e_1},P_{e_2},\cdots]^T$ is a vector of entity
authorities and $\mathcal{H}=[w^t(p_1),w^t(p_2),\cdots]^T$ is a vector
of post weights.

%% file: tracking.tex
\section{Tracking Event Evolution}\label{s:tracking}

In this section, we discuss the evolution of sketch graph and events
and develop primitive operations, which form the theoretical basis for 
our algorithms in Section \ref{s:alg}. We introduce the \emph{fading time window}
to serve as a monitor on social streams. Formally, the problem we try to solve in this paper is
shown below.

\begin{problem}
Given an evolving post network sequence
$\mathcal{G}=\{\mathcal{G}_1,\mathcal{G}_2,\cdots\}$, the event evolution tracking problem is to
generate an event set $\widetilde{S}_i$ at each moment $i$, and discover all the evolution
behaviors between events in $\widetilde{S}_i$ and
$\widetilde{S}_{i+1}$,  $\;i=1,2,\cdots$.
\end{problem}

\subsection{Incremental Tracking Framework}\label{itf}

The traditional approach taken for tracking evolving network related problems [4, 16] is illustrated
in Figure \ref{timewindow}(a). Given a post network $\mathcal{G}_t$ at time $t$, identify the event
set $\widetilde{S}_t$ associated with $\mathcal{G}_t$ (marked as step \numcircledmod{1} in the
figure). The network evolves from $\mathcal{G}_t$ to $\mathcal{G}_{t+1}$ at the next moment (step
\numcircledmod{2}). Again, the events associated with $\mathcal{G}_{t+1}$ are identified from
scratch (step \numcircledmod{3}).
Finally, the correspondence between event sets $\widetilde{S}_t$ and $\widetilde{S}_{t+1}$ is determined to
extract the evolution behaviors of events from time $t$ to $t+1$ (step \numcircledmod{4}). This
traditional approach has two major shortcomings. Firstly, repeated extraction of events from post networks from scratch is an expensive
operation. Similarly, tracing the correspondence between event sets at successive moments is also
expensive. Secondly, this step of tracing correspondence, since it is done after the two event sets
are generated, may lead to loss of accuracy. The method we propose is incremental tracking of event
evolution. It corresponds to step \numcircledmod{5} in the figure.
More precisely, for the very first snapshot of the post network, say $\mathcal{G}_0$, our approach
will generate the corresponding event set $\widetilde{S}_0$. After this, this step is never applied
again. In the steady state, we apply step \numcircledmod{5}, i.e., we incrementally derive
$\widetilde{S}_{t+1}$ from $\widetilde{S}_t$. Our experiments show that our incremental tracking
approach outperforms the traditional baselines both in quality and in performance.

\subsection{Evolution Operators}\label{s:operators}

We analyze the evolutionary process of events at each moment and
abstract them into five primitive operators: $+$, $-$, $\odot$,
$\uparrow$, $\downarrow$. We classify the operators based on the
objects they manipulate: nodes or clusters. Note that both
events and outliers are clusters.

\comment{Primitive operators can be associated with
different types of objects to express specific meanings. In
particular, we distinguish the primitive operators on post network and
clusters. Since events and outliers are mapped from clusters, they
share very similar primitives with clusters. These primitive operators
are defined below.} 
The following operators manipulate nodes in a post network. 

\vspace{-0.2cm}\begin{definition}\label{primitives:nodes}\textup{({Node Operations})}.
\squishlisttight 
\item 
$\mathcal{G}_t+p$: add a new post $p$ into
$\mathcal{G}_t(\mathcal{V}_t,\mathcal{E}_t)$ where
$p\not\in\mathcal{V}_t$. All the new edges associated with $p$ will be
constructed automatically by linkage search;

\item $\mathcal{G}_t-p$: delete a post $p$ from
$\mathcal{G}_t(\mathcal{V}_t,\mathcal{E}_t)$ where
$p\in\mathcal{V}_t$. All the existing edges associated with $p$ will
be automatically removed from $\mathcal{E}_t$.

\item $\odot (\mathcal{G}_t)$: update the weight of posts in $\mathcal{G}_t$.
\squishend 
\vspace{-0.1cm}\end{definition} 

Typically, the adding/deleting of a post will trigger the updating of
post weights. For convenience, we define two composite operators on post
networks.

\vspace{-0.2cm}\begin{definition}\label{primitives:macros}\textup{({Composite Operators})}.
\squishlisttight 
\item $\mathcal{G}_t\oplus p=\odot (\mathcal{G}_t+p)$: add a post
$p$ into $\mathcal{G}_t(\mathcal{V}_t,\mathcal{E}_t)$ where
$p\not\in\mathcal{V}_t$ and update the weight of posts in
$\mathcal{V}_t$;

\item $\mathcal{G}_t\ominus p=\odot (\mathcal{G}_t-p)$: delete a
post $p$ from $\mathcal{G}_t(\mathcal{V}_t,\mathcal{E}_t)$ where
$p\in\mathcal{V}_t$ and update the weight of posts in $\mathcal{V}_t$.
\squishend 
\vspace{-0.1cm}\end{definition} 

The following operators manipulate clusters:

\vspace{-0.2cm}\begin{definition}\label{primitives:clusters}\textup{({Cluster Operations})}.
\squishlisttight 
\item $S_t+C$: 
add cluster $C$ to the cluster set $S_t$;

\item $S_t-C$: remove cluster $C$ from the cluster set $S_t$;

\item $\uparrow (C,p)$: increase the size of $C$ by adding a new post $p$;

\item $\downarrow (C,p)$: decrease the size of $C$ by removing an old post $p$.
\squishend 
\vspace{-0.1cm}\end{definition}

Each operator defined above on a single object can be
extended to a set of objects, i.e., for a node set
$\mathcal{X}=\{p_1,p_2,\cdots,p_n\}$,
$\mathcal{G}_t+\mathcal{X}=\mathcal{G}_t+p_1+p_2+\cdots +p_n$. This 
is well defined since $+$ is associative and commutative. 

We use the left-associative convention for `$-$': that is, we write 
$A-B-C$ means $(A-B)-C$. In particular, we write 
$A-B+C$ to mean $(A-B)+C$ and the order in
which posts in a set are added/deleted to $\mathcal{G}_t$ does not matter.
These operators will be used
later in the formal description of the evolution procedures.
\comment{
We illustrate the evolution operations in the process of event evolution
tracking from post network in Figure \ref{prim:a}.
}
Figure \ref{prim:a} depicts the role played by the primitive
operators in the event evolution tracking, end to end, from the
post network onward.

\subsection{Fading Time Window}\label{timewindowmodel}
\begin{figure}[t]
\centering
\includegraphics[width=8.3cm]{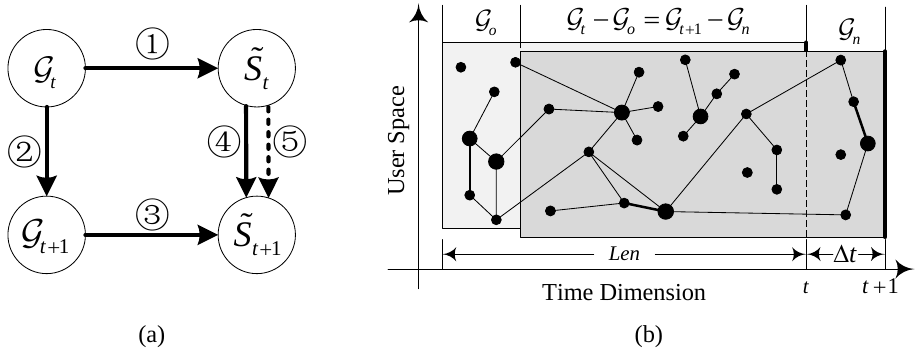}
\vspace{-0.4cm}\caption{(a) Illustration of traditional approach vs. incremental approach to
evolution tracking. Steps on edges are described in Section \ref{itf}. (b) An illustration of the
fading time window from time $t$ to $t+1$, where post weights fade w.r.t. the end of
time window. $\mathcal{G}_t$ will be updated by deleting subgraph $\mathcal{G}_{o}$ and adding
subgraph $\mathcal{G}_{n}$.}\label{timewindow}
\vspace{-0.3cm}\end{figure}

Fading (or decay) and sliding time window  are two common
aggregation schemes used in time-evolving graphs \cite{sdm06-CaoEQZ}.
Fading scheme puts a higher emphasis on newer posts, and this
characteristic has been captured by fading similarity in Eq.
(\ref{fadingsim}). Sliding time  window scheme (posts are first-in,
first-out) is essential because older posts are less important and it
is not necessary to retain all the posts in the history. Since events
evolve quickly, even inside a given time window, it is important to 
highlight new posts and degrade old posts using the fading scheme. Thus,
we combine these two schemes and introduce a \emph{fading time window},
as illustrated in Figure \ref{timewindow}(b). {In practice, users can specify the length of
the time window to adjust the scope of monitoring. Users can also choose different fading functions
to penalize old posts and highlight new posts in different ways.} Let $\Delta t$ denote the
interval between moments. For simplicity, we abbreviate the moment
$(t+i\cdot\Delta t)$ as $(t+i)$.
When the time window slides from moment
$t$ to $t+1$, the post network
$\mathcal{G}_t(\mathcal{V}_t,\mathcal{E}_t)$ will be updated to be
$\mathcal{G}_{t+1}(\mathcal{V}_{t+1},\mathcal{E}_{t+1})$. Suppose
$\mathcal{G}_{o}(\mathcal{V}_{o},\mathcal{E}_{o})$ is the old subgraph
(of $\mathcal{G}_t$)
that lapses at moment $t+1$ and
$\mathcal{G}_{n}(\mathcal{V}_{n},\mathcal{E}_{n})$ is the new subgraph
(of $\mathcal{G}_{t+1}$) that appears (see Figure \ref{timewindow}(b)). Clearly,

\begin{small}
\begin{equation}\label{windowgraph}
\mathcal{G}_{t+1}=\mathcal{G}_{t}-\mathcal{G}_{o}+\mathcal{G}_{n}
\end{equation}\vspace{-0.3cm}
\end{small}

Let $Len$ be the length of the time window. We assume
$Len>2\Delta t$, which will make
$\mathcal{V}_{o}\cap\mathcal{V}_{n}=\emptyset$. This assumption
usually holds in real applications, e.g., we set $Len$ to 1 week and
$\Delta t$ to 1 day. The following claim shows different ways to compute the overlapping part in
$\mathcal{G}_{t+1}$ and $\mathcal{G}_{t}$.
\begin{claim}\label{c:postnetwork}
\begin{small}
$\mathcal{G}_{t+1}-\mathcal{G}_{n}=\mathcal{G}_{t}-\mathcal{G}_{o}
=\mathcal{G}_{t+1}\ominus\mathcal{G}_{n}=\mathcal{G}_{t}\ominus\mathcal{G}_{o}$
\end{small}
\end{claim}
{
\proofsketch
From Eq. (\ref{windowgraph}) we know
$\mathcal{G}_{t+1}-\mathcal{G}_{n}=\mathcal{G}_{t}-\mathcal{G}_{o}$.
At the post network level, post weight is orthogonal to
node/edge sets, so we have
$\mathcal{G}_{t}\ominus\mathcal{G}_{o}=\mathcal{G}_{t}-\mathcal{G}_{o}$
and $\mathcal{G}_{t+1}\ominus\mathcal{G}_{n}=\mathcal{G}_{t+1}-\mathcal{G}_{n}$.\eop}

\subsection{Sketch Graph Evolution}\label{s:sketch}

The updating of sketch graphs from $G_t$ to $G_{t+1}$ is the core
task in event evolution tracking. We already know from Claim
\ref{c:postnetwork} that $\mathcal{G}_{t+1}-\mathcal{G}_{n}=
\mathcal{G}_{t}-\mathcal{G}_{o}$ at the post network level. This is the
overlapping part between $\mathcal{G}_{t}$ and $\mathcal{G}_{t+1}$. However,
at the sketch graph level, $Ske(\mathcal{G}_{t+1}-\mathcal{G}_{n})\neq
Ske(\mathcal{G}_{t}-\mathcal{G}_{o})$: some core posts in
$\mathcal{G}_{t}-\mathcal{G}_{o}$ may no longer be core posts due to
the removal of edges incident with nodes in $\mathcal{G}_{o}$ or simply due to
the passing of time; some
non-core posts may become core posts because of the adding of edges
with nodes in $\mathcal{G}_{n}$. To measure the changes in the
overlapping part, we define the following three components.

\vspace{-0.1cm}\begin{definition}\label{d:components} \textup{({Updated Components in
Overlap})}.
\squishlisttight 
\item $\overline{S_{+}}$: components of non-core posts in
$\mathcal{G}_{t}-\mathcal{G}_{o}$ that become core posts in
$\mathcal{G}_{t+1}-\mathcal{G}_{n}$ due to the adding of
$\mathcal{G}_{n}$, i.e.,

 $\overline{S_{+}}=Ske(\mathcal{G}_{t+1}-\mathcal{G}_{n})-Ske(\mathcal{G}_{t+1}\ominus\mathcal{G}_{n})$

\item $\overline{S_{-}}$: components of core posts in
$\mathcal{G}_{t}-\mathcal{G}_{o}$ that become non-core posts in
$\mathcal{G}_{t+1}-\mathcal{G}_{n}$ due to the removing of
$\mathcal{G}_{o}$, i.e.,

 $\overline{S_{-}}=Ske(\mathcal{G}_{t}-\mathcal{G}_{o})-Ske(\mathcal{G}_{t}\ominus\mathcal{G}_{o})$

\item $\overline{S_{\odot}}$: components of core posts in
$\mathcal{G}_{t}-\mathcal{G}_{o}$ that become non-core posts in
$\mathcal{G}_{t+1}-\mathcal{G}_{n}$ due to the passing of time, i.e.,

$\overline{S_{\odot}}=Ske(\mathcal{G}_{t}\ominus\mathcal{G}_{o})-Ske(\mathcal{G}_{t+1}\ominus\mathcal{G}_{n})$
\squishend 
\end{definition}

Based on Definition \ref{d:components}, Theorem \ref{t:overlap} shows
how the overlapping parts in $\mathcal{G}_t$ and $\mathcal{G}_{t+1}$ differ at the sketch
graph level.
\from{laks}{I changed $G$'s to $\mathcal{G}$'s above: pl. check!}

\begin{theorem}\label{t:overlap}
From moment $t$ to $t+1$, the changes of core posts in the overlapping
part, i.e., $\mathcal{G}_{t+1}-\mathcal{G}_{n}$ (equivalently,
$\mathcal{G}_{t}-\mathcal{G}_{o}$), can be updated using the components
$\overline{S_{+}}$, $\overline{S_{-}}$ and $\overline{S_{\odot}}$.
That is,
{\small
\begin{equation}\label{e:overlap}
Ske(\mathcal{G}_{t+1}-\mathcal{G}_{n})-Ske(\mathcal{G}_{t}-\mathcal{G}_{o})=
\overline{S_{+}}-\overline{S_{-}}-\overline{S_{\odot}}
\end{equation}}
\end{theorem}
{\proofsketch $\overline{S_{+}}-\overline{S_{-}}-\overline{S_{\odot}}=Ske(\mathcal{G}_{t+1}-\mathcal{G}_{n})
-Ske(\mathcal{G}_{t+1}\ominus\mathcal{G}_{n})-(Ske(\mathcal{G}_{t}-\mathcal{G}_{o})-
Ske(\mathcal{G}_{t}\ominus\mathcal{G}_{o}))-(Ske(\mathcal{G}_{t}\ominus\mathcal{G}_{o})-Ske(\mathcal{G}_{t+1}\ominus\mathcal{G}_{n}))=
Ske(\mathcal{G}_{t+1}-\mathcal{G}_{n})-Ske(\mathcal{G}_{t}-\mathcal{G}_{o})$.
\eop}

\from{laks}{The use of minus in the above proof is ambiguous. Please use paretheses to
clearly indicate grouping. Also, the proof is dry: you should say something about how you go from the
first equality to the second equality. Are you canceling terms? How?}

The following theorem shows the iterative and incremental updating of
sketch graphs from moment $t$ to $t+1$.

\begin{theorem}\label{t:sketchevo}
From moment $t$ to $t+1$, the sketch graph evolves by removing core
posts in $\mathcal{G}_{o}$, adding core posts in $\mathcal{G}_{n}$ and
updating core posts in the overlapping part. That is
{\begin{equation}\label{e:sketchevo}
\overline{S_{t+1}}= {G}_{t+1} =
\overline{S_{t}}-\overline{S_{o}}-\overline{S_{-}}-\overline{S_{\odot}}+\overline{S_{n}}+\overline{S_{+}}
\end{equation}}
\end{theorem}

{\proofsketch $\overline{S_{+}}-\overline{S_{-}}-\overline{S_{\odot}}=Ske(\mathcal{G}_{t+1}-\mathcal{G}_{n})
-Ske(\mathcal{G}_{t+1}\ominus\mathcal{G}_{n})-(Ske(\mathcal{G}_{t}-\mathcal{G}_{o})-
Ske(\mathcal{G}_{t}\ominus\mathcal{G}_{o}))-(Ske(\mathcal{G}_{t}\ominus\mathcal{G}_{o})-Ske(\mathcal{G}_{t+1}\ominus\mathcal{G}_{n}))=
Ske(\mathcal{G}_{t+1}-\mathcal{G}_{n})-Ske(\mathcal{G}_{t}-\mathcal{G}_{o})$.
\eop}

Theorem \ref{t:sketchevo} indicates that we can incrementally maintain $\overline{S_{t+1}}$ from
$\overline{S_{t}}$. Since we define events based on connected components in a sketch graph, this
incremental updating of sketch graphs benefits incremental computation of event evolution
essentially.

\from{laks}{Again, use parentheses to indicate in what order the opetations, esp. the minuses,
are applied. Also, please some explanation to make the proof more intuitive and less dry.}

\subsection{Event Evolution}\label{s:event}

\begin{figure}[t]
\centering
 \subfigure[]{
\label{prim:a}
{\noindent\includegraphics[width=8cm]{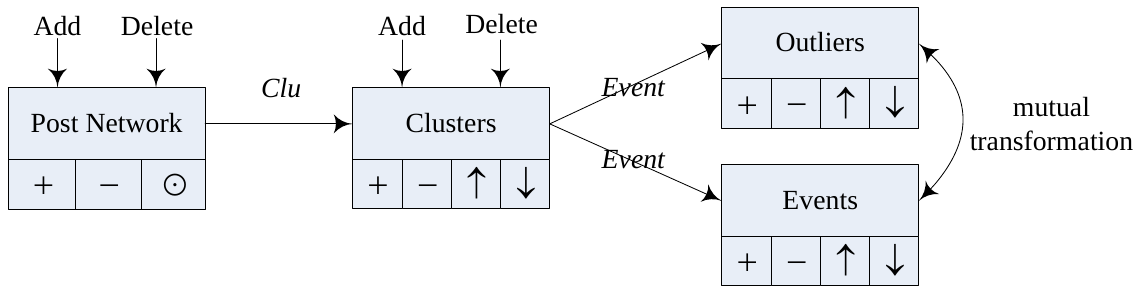}}}
\subfigure[]{
\label{prim:b}
{\small
\begin{tabular}{|c|c|c|c|}
\hline
$|N_c(p)|$ & 0 & 1 & $\geq 2$\\\hline
Add a core post $p$ & $+$ & $\uparrow$ & $Merge$\\\hline
Delete a core post $p$ & $-$ & $\downarrow$ & $Split$\\\hline
\end{tabular}}}
\vspace{-0.4cm}\caption{(a) The relationships between primitives and evolutions.
Each box represents an evolution object and the arrows between them
describe inputs/outputs. (b) The evolutionary behavior table for
clusters when adding or deleting a core post $p$.}
 \label{sc}
\vspace{-0.3cm}\end{figure}

Let $Clu(\mathcal{G}_t)$
denote the clustering operation on post network $\mathcal{G}_t$. The clustering
will shift and reorder the nodes in $\mathcal{V}_t$ by grouping nodes
that belong to the same cluster together. We use
$S_t=Clu(\mathcal{G}_t)$ to represent the set of clusters obtained
from $\mathcal{G}_t$. Notice that noise posts in $\mathcal{G}_t$ do
not appear in any cluster, so the number of posts in $S_t$ is typically smaller
than $|\mathcal{V}_t|$.
Similarly, we let $Clu(\mathcal{G}_{n})=S_{n}$ and $Clu(\mathcal{G}_{o})=S_{o}$ be the
cluster sets on the graphs $\mathcal{G}_{n}$ and $\mathcal{G}_{o}$.

\stitle{Cluster Evolution}. Based on Definition \ref{d:cluster}, a
cluster at moment $t$ is generated from a component in
$\overline{S_{t}}$. We can apply the $Gen$ function on sketch graphs to get the iterative
computation of clusters.

\begin{corollary}\label{t:clusterevo}
From moment $t$ to $t+1$, the set of clusters can be incrementally
updated by the iterative computation
{\small
\vspace{-0.2cm}\begin{equation}\label{e:clusterevo}
{S_{t+1}} = Clu(\mathcal{G}_{t+1})=
{S_{t}}-{S_{o}}-{S_{-}}-{S_{\odot}}+{S_{n}}+{S_{+}}
\end{equation}}
\vspace{-0.4cm}\end{corollary}
\vspace{-0.3cm}{\proofsketch
Since the generation of any two clusters from two different components
is independent, by applying $Gen$ function on both sides of Eq.
(\ref{e:sketchevo}) we get the iterative computation of
clusters.\eop}

The basic operations underlying Eq. (\ref{e:clusterevo}) are the cases when
$S_t$ is modified by the addition or deletion of a cluster that includes
only one post. In the following, we analyze and show the evolution of clusters
by adding a post $p_n$ or deleting a post $p_o$. Let
$N_c(p_n)$ denote the set of clusters where $p_n$'s neighboring core
posts belong {\sl before} $p_n$ is {\sl added}, and let $N_c(p_o)$ denote the set
of clusters where $p_o$'s neighboring core posts belong {\sl after} $p_o$ is
{\sl removed}. It should be noted that $|N_c(p)|=0$ means $p$ has no
neighboring core posts. Notice that $Merge$ and $Split$ are
composite operations and can be composed by a series of cluster
primitive operations. The evolution behaviors of clusters is shown in
Figure \ref{prim:b} and explained below. 

\from{laks}{I think we agreed the operations $+, -$ etc should be
binary. Please make sure both operands are mentioned below.}

{
\line(1,0){230}

\textbf{(a) Addition}: $S_t+\{p_n\}$

 If $p_n$ is a noise post after being added into $\mathcal{G}_t$,
ignore $p_n$. If $p_n$ is a border post, add $p_n$ to each cluster in
$N_c(p_n)$. Else, $p_n$ is a core post and we do the following:
\squishlisttight 
\item If $|N_c(p_n)|=0$: $+C$ and $C=\{p_n\}\cup\mathcal{N}(p_n)$;

\item If $|N_c(p_n)|=1$: $\uparrow (C,\{p_n\}\cup\mathcal{N}(p_n))$;

\item If $|N_c(p_n)|\geq 2$: $Merge=+C-\sum_{C'\in N_c(p_n)}{C'}$
and $C=N_c(p_n)\cup \{p_n\}\cup\mathcal{N}(p_n)$.
\squishend 
\vspace{0.2cm}\textbf{(b) Deletion}: $S_t-\{p_o\}$

 If $p_o$ is a noise post before being deleted from $\mathcal{G}_t$,
ignore $p_o$. If $p_o$ is a border post, delete $p_o$ from each
cluster in $N_c(p_o)$. Else, $p_o$ is a core post and we do the
following:
\squishlisttight
\item If $|N_c(p_o)|=0$: $-C$ where $p_o\in C$;

\item If $|N_c(p_o)|=1$: $\downarrow (C,\{p_o\}\cup\mathcal{N}(p_o))$;

\item If $|N_c(p_o)|\geq 2$: $Split=-C+\sum_{C'\in N_c(p_o)}{C'}$,
$p_o\in C$.
\squishend

\line(1,0){230}}

Using commutativity and associativity of operator `$+$' and 
the left-associative convention for operator `$-$',
we can rewrite the expression
$S_t-p_1+p_2-p_3$ as $S_t-(p_3+p_1)+p_2$.
While both expressions are equivalent, in terms of the total number of primitive cluster operations,
different ordering of cluster sequences may have different
performance. In the following, we show a theorem to indicate how to reduce the
number of primitive cluster operations by reordering. As an example, $S_t+p_1-p_2+p_3-p_4$ 
can be reordered as $S_t-(p_2+p_4)+(p_1+p_3)$ to reduce the number of primitive operations 
during cluster evolution.

\begin{theorem}\label{sequence}
Suppose posts have an equal probability to join any cluster. Given an initial set of clusters and an
arbitrary sequence of node addition and deletions, the number of
cluster primitive operations can be reduced by performing all the
deletions first.  

\end{theorem}

{\proofsketch
Since posts have equal probability to join clusters, a smaller $|\mathcal{N}(p)|$
(i.e., the \#edges of $p$) indicates a smaller number of neighboring clusters.
Clearly, if the deletions are performed first, we can save time because there will be no edges
between the deleted nodes and added nodes. Otherwise, the edges between the
deleted nodes and added nodes need to be constructed first and
removed later, which implies a higher $|\mathcal{N}(p)|$ in computation.\eop}

\stitle{The Evolution of Events and Outliers}. Recall, at any moment, an outlier may grow
into an event and an event may degrade into an outlier. 
Based on Definition \ref{d:event}, the evolution procedures of
events and outliers follow the evolution of clusters.
That is, the addition of a new event or outlier as well as the deletion of
an existing event or outlier is exactly the same as for a cluster. The only difference
is that the $Event$ function is applied to check whether the cluster added/deleted
is an event or an outlier. When the size of a cluster increases (as with $\uparrow (C,p)$)
it may correspond to an outlier/event growing into a larger outlier/event,
or to an outlier growing in size, exceeding the threshold $\varphi$ and turning into an event.
Similarly, when a cluster shrinks (as with $\downarrow (C,p)$) it may be an outlier/event becoming
a smaller outlier/event or an event becoming smaller than the threshold
and turning into an outlier. 

%% file: alg.tex
\section{Incremental Algorithms}\label{s:alg}

The approach of decomposing an evolving graph
into a series of snapshots  for each moment is a traditional way to
tackle evolving graph related problems \cite{kdd07-AsurPU,pvldb09-KimH}.
However, this traditional approach suffers from both quality and performance, since
events are generated from scratch and are matched heuristically. To overcome these challenges, we
propose an incremental tracking framework, as introduced in Section \ref{itf} and illustrated
in Figure \ref{timewindow}(a).

In this section, we leverage our primitive operators for evolution tracking by
proposing Algorithms \ctrack and \etrack to track the evolution of clusters and
events respectively. The observation is that at each moment
$|\mathcal{V}_{o}|+|\mathcal{V}_{n}|\ll |\mathcal{V}_t|$, where $\mathcal{V}_o$ and
$\mathcal{V}_{n}$ are the set of old and new posts between moments $t$ and $t+1$.
So we can
save a lot of computation by adjusting clusters and events
incrementally, rather than generating them from scratch.

\stitle{Bulk Updating}. Traditional incremental computation on
evolving graphs usually treats the addition/deletion of nodes or edges
one by one \cite{kdd07-ChenT,tkdd09-WanNDYZ}. In Section
\ref{s:event}, we  discussed the updating of $S_t$ by adding or
deleting a single post. However, in a real scenario, since social posts
arrive at a  high speed, the post-by-post incremental updating
will lead to poor performance. In this section, we speed up the incremental
computation of $S_t$ by {\em bulk updating}. We define a bulk as a cluster
of posts and ``lift'' the one-by-one
updating of $S_t$ to bulk updating. Recall that $S_n$ and $S_o$
denote the set of clusters in the graph $\mathcal{G}_n = \mathcal{G}_{t+1} - \mathcal{G}_{t}$
 and $\mathcal{G}_o = \mathcal{G}_{t} - \mathcal{G}_{t+1}$ respectively.
Specifically, given a cluster $C_n\in S_{n}$, let
$N_c(\overline{C_n})$ denote the set of clusters in $\mathcal{G}_t$
where neighboring core posts of nodes in $\overline{C_n}$ belong,
i.e., $N_c(\overline{C_n})=\cup_{p_n\in\overline{C_n}}{N_c(p_n)}$.
Analogously, given $C_o\in S_{o}$, we let
$N_c(\overline{C_o})=\cup_{p_o\in\overline{C_o}}{N_c(p_o)}$.
Clearly, updating $S_t$ with a single node $\{p\}$ is a special case of bulk updating.

\stitle{cTrack}. The steps for incremental tracking of cluster
evolution (\ctrack) are summarized in Algorithm \ref{alg1}. \ctrack
follows the iterative computation in Eq. (\ref{e:clusterevo}) and sequence order in 
Theorem \ref{sequence}, that is
${S_{t+1}} = {S_{t}}-{S_{o}}-{S_{-}}-{S_{\odot}}+{S_{n}}+{S_{+}}$. 
As analyzed in Section \ref{s:event}, each bulk addition and bulk deletion has three possible evolutionary
behaviors, decided by the size of $N_c(\overline{C})$. {Lines 3-13 deal with deleting a
bulk $C_o$, where \{$-$, $\downarrow$, $Split$\} patterns are handled. Lines 15-27 deal with
adding a bulk $C_n$ and handle \{$+$, $\uparrow$, $Merge$\} patterns.}
The time complexity of \ctrack is linear in the total number of bulk
updates.

\begin{algorithm}[t]
\caption{\ctrack}
\label{alg1}
\begin{small}
\KwIn{$S_t$, ${S}_{o}$, ${S}_{n}$, $S_{-}$, $S_{+}$, $S_{\odot}$}
\KwOut{$S_{t+1}$}
$S_{t+1}=S_t$\;
\tcp{Delete ${S}_{o}\cup S_{-}$}
\For{each cluster $C_o$ in ${S}_{o}\cup S_{-}\cup S_{\odot}$}
{
   $\overline{C_o}=Ske(C_o)$\;
   $N_c(\overline{C_o})=\cup_{p_o\in\overline{C_o}}{N_c(p_o)}$\;
   \If{$|N_c(\overline{C_o})|=0$}
   {
       remove cluster $C_o$ from $S_{t+1}$\;
   }
   \ElseIf{$|N_c(\overline{C_o})|=1$}
   {
       delete $C_o$ from cluster $C'$ where $C'\in N_c(\overline{C_o})$\;
   }
   \Else
   {
       remove the cluster that $C_o$ belongs to from $S_{t+1}$\;
       \For{each cluster $C'\in N_c(\overline{C_o})$}
       {
           assign a new cluster id for $C'$\;
           add $C'$ into $S_{t+1}$\;
       }
   }
}
\tcp{Add ${S}_{n}\cup S_{+}$}
\For{each cluster $C_n$ in ${S}_{n}\cup S_{+}$}
{
   $\overline{C_n}=Ske(C_n)$\;
   $N_c(\overline{C_n})=\cup_{p_n\in\overline{C_n}}{N_c(p_n)}$\;
   \If{$|N_c(\overline{C_n})|=0$}
   {
       assign a new cluster id for $C_n$ and add $C$ to $S_{t+1}$\;
   }
   \ElseIf{$|N_c(\overline{C_n})|=1$}
   {
       add $C_n$ into cluster $C'$ where $C'\in N_c(\overline{C_n})$\;
   }
   \Else
   {
       assign a new cluster id for $C_{n}$\;
       \For{each cluster $C'\in N_c(\overline{C_n})$}
       {
           $C_{n}=C_{n}\cup C'$\;
           remove $C'$ from $S_{t+1}$\;
       }
       $C_{n}=C_{n}\cup C_n$\;
       add $C_{n}$ into $S_{t+1}$\;
   }
}
return $S_{t+1}$\;
\end{small}
\end{algorithm}

\stitle{eTrack}. Algorithm \etrack works on top of \ctrack. We summarize the
steps for incremental event tracking (\etrack) in Algorithm
\ref{alg2}. Based on \ctrack, \etrack monitors the changes in the
cluster set effected by \ctrack at each moment. If the cluster is not
changed, \etrack will take no action; otherwise, \etrack will judge
the corresponding cases and invoke $Event$ function to handle the event
evolution behaviors (Lines 4-13). \from{laks}{Map versus Event issue.} Notice that in Lines 5-9,
if a cluster $C$ in $S_i$ has ClusterId $id$, we use the convention that $S_i(id)=C$, and
$S_i(id)=\emptyset$ means there is no cluster in $S_i$ with ClusterId
$id$. {Especially, lines 7-9 mean an event in $S_i$ evolves into an event in $S_{i+1}$
by deleting the part $S_i(id)-S_{i+1}(id)$ first and adding the part $S_{i+1}(id)-S_{i}(id)$
later.}

\begin{algorithm}[t]
\caption{\etrack}
\label{alg2}
{\small
\KwIn{$\mathcal{G}=\{\mathcal{G}_1,\mathcal{G}_2,\cdots,\mathcal{G}_n\}$, $S_1$}
\KwOut{Event evolution behaviors}
\For{$i$ from 1 to $n-1$}
{
   obtain ${S}_{o}$, ${S}_{n}$, $S_{-}$, $S_{+}$ from
$\mathcal{G}_{i+1}-\mathcal{G}_i$\;
   $S_{i+1}=$\ctrack$(S_i,{S}_{o},{S}_{n},S_{-},S_{+},S_{\odot})$\;
   \For{each cluster $C\in S_{i+1}$}
   {
       $id$ = $ClusterId(C)$\;
       \If{$S_i(id)\neq\emptyset$}
       {
           output $\downarrow (C,S_i(id)-S_{i+1}(id))$\;
           output $\uparrow (C,S_{i+1}(id)-S_{i}(id))$\;
           $Event(C)$\;
       }
       \lElse{$Event(+C)$\;}
   }
   \For{each cluster $C\in S_{i}$}
   {
        $id$ = $ClusterId(C)$\;
        \lIf{$S_{i+1}(id)=\emptyset$}{$Event(-C)$\;}
   }
}
}
\end{algorithm}

%% file: exp.tex
\section{Experiments}\label{s:exp}

In this section, we first ``tune'' the construction of post network and sketch graph to
find the best selection of fading similarity measures and density
parameters. Then, we test the quality and performance of event
evolution tracking algorithms on two social streams: Tech-Lite and
Tech-Full that we crawled from Twitter. 
{We have designed two types of baseline algorithms for event detection
and evolution tracking. Our event detection baseline covers the major techniques reported in
\cite{kdd09-LeskovecBK,chi11-MarcusBBKMM, www10-SakakiOM,wsdm11-SarmaJY}. Our tracking baseline
summarizes the state of the art algorithms reported in \cite{kdd07-AsurPU,pvldb09-KimH}.}
All experiments are conducted on a computer with Intel 2.66
GHz CPU, 4 GB RAM, running 64-bit Windows 7. All algorithms are
implemented in Java. We use a graph database called
Neo4J\footnote{http://neo4j.org/} to store and manipulate the post network and sketch
graph.

\begin{figure}[t]
\centering
\includegraphics[width=6.5cm]{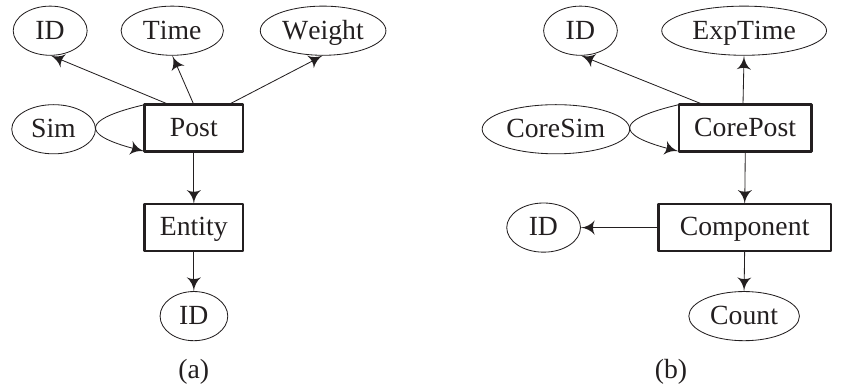}
\vspace{-0.2cm}\caption{Graph schemas of (a) post network and (b) sketch graph.
Rectangles represent nodes and ellipses represent attributes. ``Sim''
means similarity edges between posts and ``CoreSim'' means core
edges. ``Weight'' is the post weight. ``ExpTime'' is the time
when a core post expires due to time passing. ``Count'' is the number of
core posts in a component. }\label{graphschema}
\vspace{-0.5cm}\end{figure}
\begin{figure}[t]
\centering
\includegraphics[width=8cm]{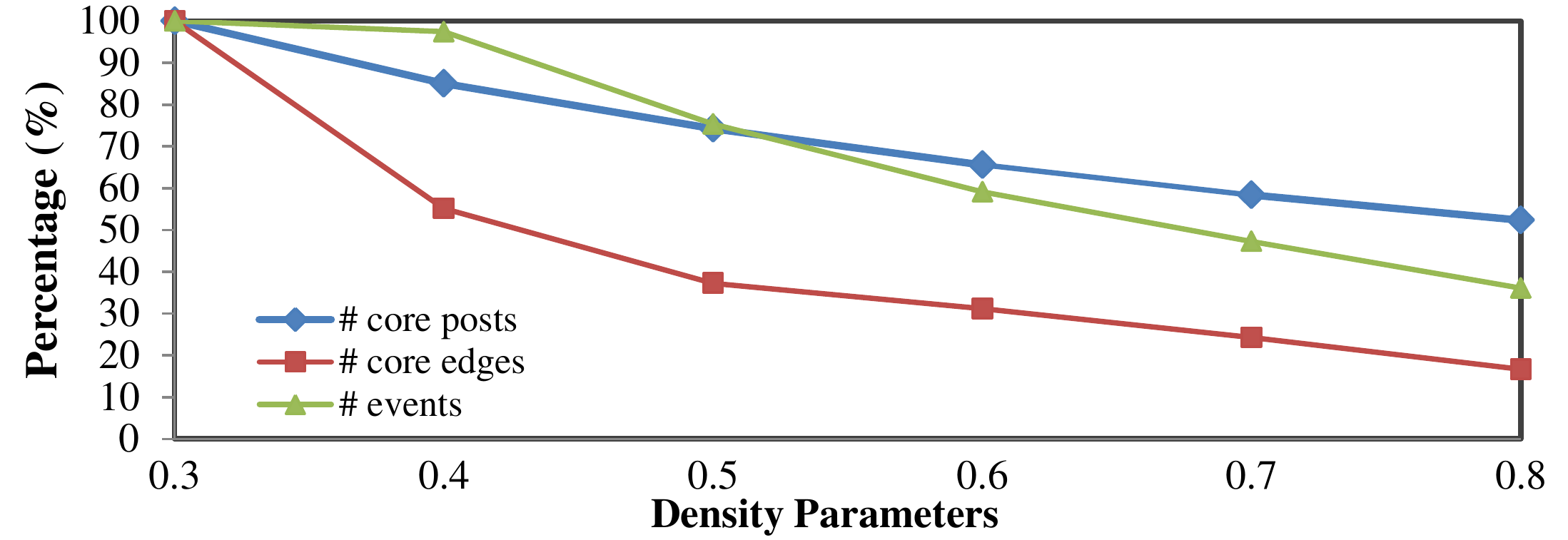}
\vspace{-0.2cm}\caption{The changing trends of the number of core posts, core edges and events when
increasing $\delta_1=\varepsilon_1$ from 0.3 to 0.8. We set $\delta_1=\varepsilon_1=0.3$ as the 100\% basis and keep $\varepsilon_0=0.2$.
}\label{density}\vspace{-0.2cm}
\end{figure}

\stitle{Datasets}. All datasets are crawled from
Twitter.com via Twitter API. Although our event tracking algorithm
works regardless of the domain, we make the
data set domain specific in order to facilitate evaluation of the generated results.
We built a technology domain
dataset called Tech-Lite by aggregating all the timelines of users
listed in Technology
category of ``Who to follow''\footnote{http://twitter.com/who\_to\_follow/interests} and their retweeted users. Tech-Lite has 352,328
tweets, 1402 users and the streaming rate is about 11700 tweets/day.
Based on the intuition that users followed by users in Technology category are
most likely also in the technology domain, we obtained a larger
technology social stream called Tech-Full by collecting all the
timelines of users that are followed by users in Technology category. Tech-Full
has 5,196,086 tweets, created by 224,242 users, whose streaming
rate is about 7216 tweets/hour. Both timelines in Tech-Lite and
Tech-Full include retweets and have a time span from Jan. 1 to Feb. 1,
2012. Since Tech-Lite is a sampled subset of Tech-Full, the parameters learned from Tech-Lite generally apply to Tech-Full.

\stitle{Graph Storage Schema}. Neo4J provides convenience in managing
a large and fast-evolving data structure that represents a post
network and sketch graph. A database in Neo4J is based on the
attributed graph model \cite{Srinivasa11}, where each node/edge can be
associated with multiple attributes in the form of key-value pairs.
Graph schemas of the post network and sketch graph are shown in Figure
\ref{graphschema} (a) and (b) respectively. Indices are created on
attributes to support fast exact queries and range queries of
nodes.

\subsection{Tuning Post Network and Sketch Graph}\label{s:tuning}

The quality of post network and sketch graph construction will determine the quality of event
generation. Recall from Section \ref{pnc} that the
following factors influence the construction of post networks and
sketch graphs: (1) entity extraction from post contents; (2)
similarity/distance measures between posts; and (3) density parameters
for the generation of core posts and core edges. We measure the
influence of each factor below to make informed choices.

\stitle{Post Content Preprocessing}. As described in Section
\ref{pnc}, we extract entities from posts by Stanford POS tagger.
\from{laks}{name the tagger and give a url.}
One alternative approach to entity extraction is using
hashtags. However, 
only 11\% of the tweets in
our dataset have hashtags. This will result in lots of posts in the dataset
having no similarity score between them. Another approach 
is simply tokenizing tweets into unigrams and treating unigrams as entities, and we call it the
``Unigrams'' approach. This approach is based on the state of art method for 
event detection as discussed in \cite{chi11-MarcusBBKMM}.
Table 2(a) shows the comparison of the 
three entity extraction approaches in the first time window of the
Tech-Full social stream.
If we use ``Unigrams'',
obviously the number of entities is larger than other two
approaches, but the number of edges between posts tends to be smaller,
because tweets written by different users usually share very few
common words even when they talk about the same event. The
``Hashtags'' approach also produces a smaller number of edges, core
posts and events, since it generates a much sparser post network.
Overall, the ``POS-Tagger'' approach can discover more similarity
relationships between posts and produce more core posts and events
given the same social stream and parameter setting.

\begin{table}[t]
\centering
{\small\noindent
\subfigure[Results of different entity extraction approaches.]{\label{tuning:a}
\begin{tabular}{|c|c|c|c|c|}
\hline
Methods & \#edges & \#coreposts & \#coreedges & \#events \\\hline
Hashtags & 182905 & 6232 & 28964 & 196 \\\hline
Unigrams & 142468 & 15070 & 46783 & 430 \\\hline
\textbf{POS-Tagger} & 357132 & 21509 & 47808 & 470 \\\hline
\end{tabular}}
\subfigure[Results of different distance functions.]{\label{tuning:b}
\begin{tabular}{|c|c|c|c|c|}
\hline
Methods & \#edges & \#coreposts & \#coreedges & \#events \\\hline
No-Fading & 1159364 & 57291 & 32875 & 510 \\\hline
Exp-Fading & 327390 & 7655 & 46075 & 148 \\\hline
\textbf{Reci-Fading} & 357132 & 21509 & 47808 & 470 \\\hline
\end{tabular}}
\subfigure[Precision and recall of top 50 events.]{\label{tuning:c}
\begin{tabular}{|c|c|c|c|}
\hline
\multirow{2}{1.5cm}{Methods} & Precision & Recall & Precision \\
 & (major events) & (major events) & (G-Trends)\\\hline
HashtagPeaks & 0.40 & 0.30 & 0.25 \\\hline
UnigramPeaks & 0.45 & 0.40 & 0.20 \\\hline
Louvain & 0.60 & 0.55 & 0.75\\\hline
\textbf{eTrack} & 0.80 & 0.80 & 0.95 \\\hline
\end{tabular}}
}
\vspace{-0.2cm}\caption{(a) and (b) are in the first time window of Tech-Lite with 75,849 posts in one
week. Density parameters
$(\delta_1,\varepsilon_0,\varepsilon_1)=(0.5,0.2,0.5)$ for core posts
and threshold $\varphi=10$ for event identification. (c) shows the
precision and recall of top 20 events generated by Baseline 1a, 1b, \louvain and
\etrack. }\label{tuning}\vspace{-0.3cm}
\end{table}

\stitle{Similarity/Distance Measures}. Many set-based similarity
measures such as Jaccard coefficient \cite{IRBook} can be used to compute the similarity $S(p_i^L,p_j^L)$
between posts. Since entity frequency is usually 1 in a tweet,
measures such as Cosine similarity and Pearson correlation
\cite{IRBook} will degenerate to a form very similar to Jaccard, so we use Jaccard.
The distance function $D(|p_i^\tau-p_j^\tau|)$ along time dimension
determines how similarity to older posts is penalized, with respect to
recent posts. We compared three
different distance functions: (1) reciprocal fading
(``Reci-Fading'') with
$D(|p_i^\tau-p_j^\tau|)={|p_i^\tau-p_j^\tau|+1}$, (2) exponential
fading (``Exp-Fading'') with
$D(|p_i^\tau-p_j^\tau|)={e^{|p_i^\tau-p_j^\tau|}}$ and (3) no fading
(``No-Fading'') with $D(|p_i^\tau-p_j^\tau|)=1$. For any posts $p_i, p_j$, clearly $e^{|p_i^\tau-p_j^\tau|}\geq{|p_i^\tau-p_j^\tau|+1}\geq
1$. Since a time window usually contains many moments,
Exp-Fading penalizes the posts in the old part of time window
severely (see Table 2(b)): the number of core posts and events generated by Exp-Fading is much
lower than by other approaches. Since No-Fading does not penalize old posts in the time window, too many edges and core posts will be
generated without considering recency. Reci-Fading is in between,
which is a 
more gradual penalty function than Exp-Fading and we use it by
default.

\stitle{Density Parameters}. The density parameters
$(\delta_1,\varepsilon_0,\varepsilon_1)$ control the construction of the
sketch graph. Clearly, the higher the density parameters, the
smaller and sparser the sketch graph. Figure \ref{density} shows the number of core posts,
core edges and events as a percentage of the numbers for $\delta_1=\varepsilon_1=0.3$,
as $\delta_1=\varepsilon_1$ increases from 0.3 to 0.8. Results are obtained from the
first time window of the Tech-Lite social stream. We can see the rate of decrease of $\#$events
is higher than the rates of $\#$core posts and $\#$core edges, because
events are less likely to form in sparser sketch graphs. More small events
can be detected by lower density parameters, but the computational
cost will increase because of larger and denser sketch graphs. However, for big
events, they are not very sensitive to these density parameters. We set
$\delta_1=\varepsilon_1=0.5$ as a trade-off between the compactness of events and complexity.

\subsection{Event Evolution Tracking}

\begin{figure}[t]
\centering
\includegraphics[width=7.5cm]{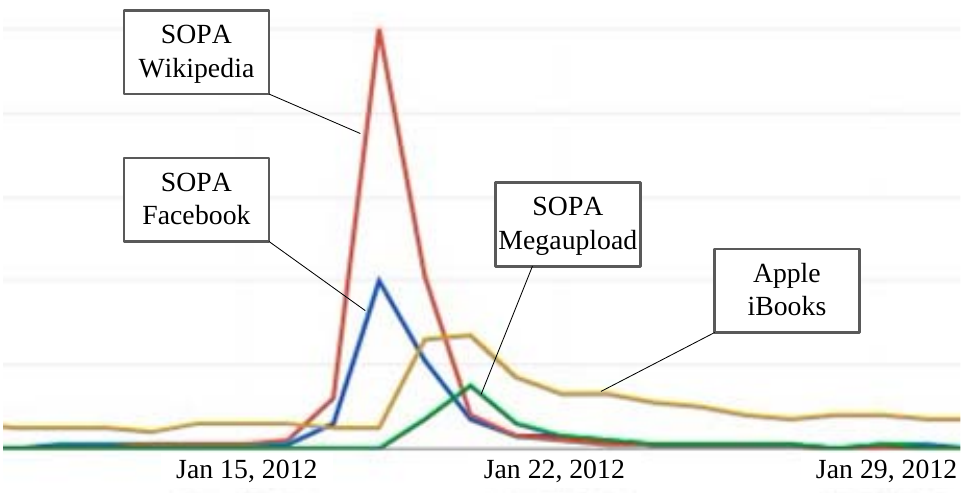}
\caption{Examples of Google Trends peaks in January 2012. We validate the events generated by cTrack
by checking the existence of volume peaks at a nearby time moment in Google Trends.
Although these peaks can detect bursty events, Google Trends cannot discover the merging/splitting patterns.}\label{googlepeaks}
\vspace{-0.2cm}
\end{figure}

\begin{figure*}[t]
\centering
\includegraphics[width=15cm]{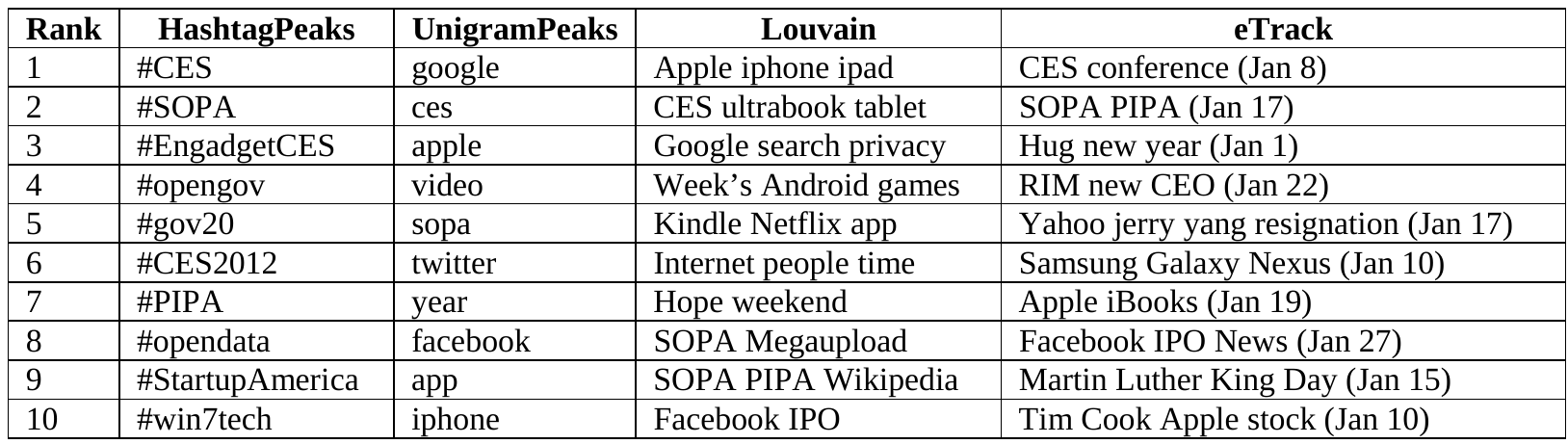}\vspace{-0.2cm}
\caption{Lists of top 10 events detected from Twitter Technology streams in January 2012
by baseline HashtagPeaks, UnigramPeaks, Louvain and our approach eTrack. To adapt for event
detection, Louvain method defines events by post communities. We annotate events generated by both
Louvain and eTrack by highly frequent entities. eTrack also shows the birthday
of events.}\label{top10}\vspace{-0.3cm}
\end{figure*}

\stitle{Ground truth}. To generate the ground truth, we crawl news articles in January 2012 from
famous technology websites such as TechCrunch, Wired, CNET, etc, without looking at tweets. Then we
treat the titles of news as posts and apply our event tracking algorithm to extract event evolution
patterns.
Finally, a total of 20 major events with life cycles are identified as
ground truth. Typical events include ``happy new year'', ``CES 2012'',
``jerry yang yahoo resignation'', ``sopa wikipedia blackout'', etc. 

To find more noticeable events in the technology news domain, we use Google Trends for
Search\footnote{http://www.google.com/trends/}, which shows the traffic trends of keywords that
appeared in Google Search along the time dimension. If an event-indicating phrase has
a volume peak in Google Trends at a specific time, it means this event is sufficiently validated
by the real world. We validate the correctness of an event $E_i$ generated by our approach by the
following process:
we pick the top 3 entities of $E_i$ ranked by frequency in $E_i$ and search them in Google Trends,
and if the traffic trend of these top entities has a distinct peak at a nearby time to $E_i$, we
consider that $E_i$ corresponds to a real world event widely witnessed by the public. Four examples
of Google Trends peaks are shown in Figure \ref{googlepeaks}.

\stitle{Baseline 1: Peak-Detection}. In recent works \cite{kdd09-LeskovecBK,chi11-MarcusBBKMM,
www10-SakakiOM,wsdm11-SarmaJY}, events are generally detected as volume peaks of phrases over time
in social streams. These approaches share the same spirit that aggregates the
 frequency of event-indicating phrases at each moment to build a histogram and generates events by
 detecting volume peaks in the histogram. We design two variants of Peak-Detection to
 capture the major techniques used by these state-of-the-art approaches.

\squishlisttight 
\item Baseline 1a: \hashtags: aggregates frequent hashtags;
 
\item Baseline 1b: \unigrams: aggregates frequent unigrams. 
\squishend

\begin{figure*}[t]
\centering
\subfigure[The life cycle of ``CES Conference'']{
  \includegraphics[height=6cm] {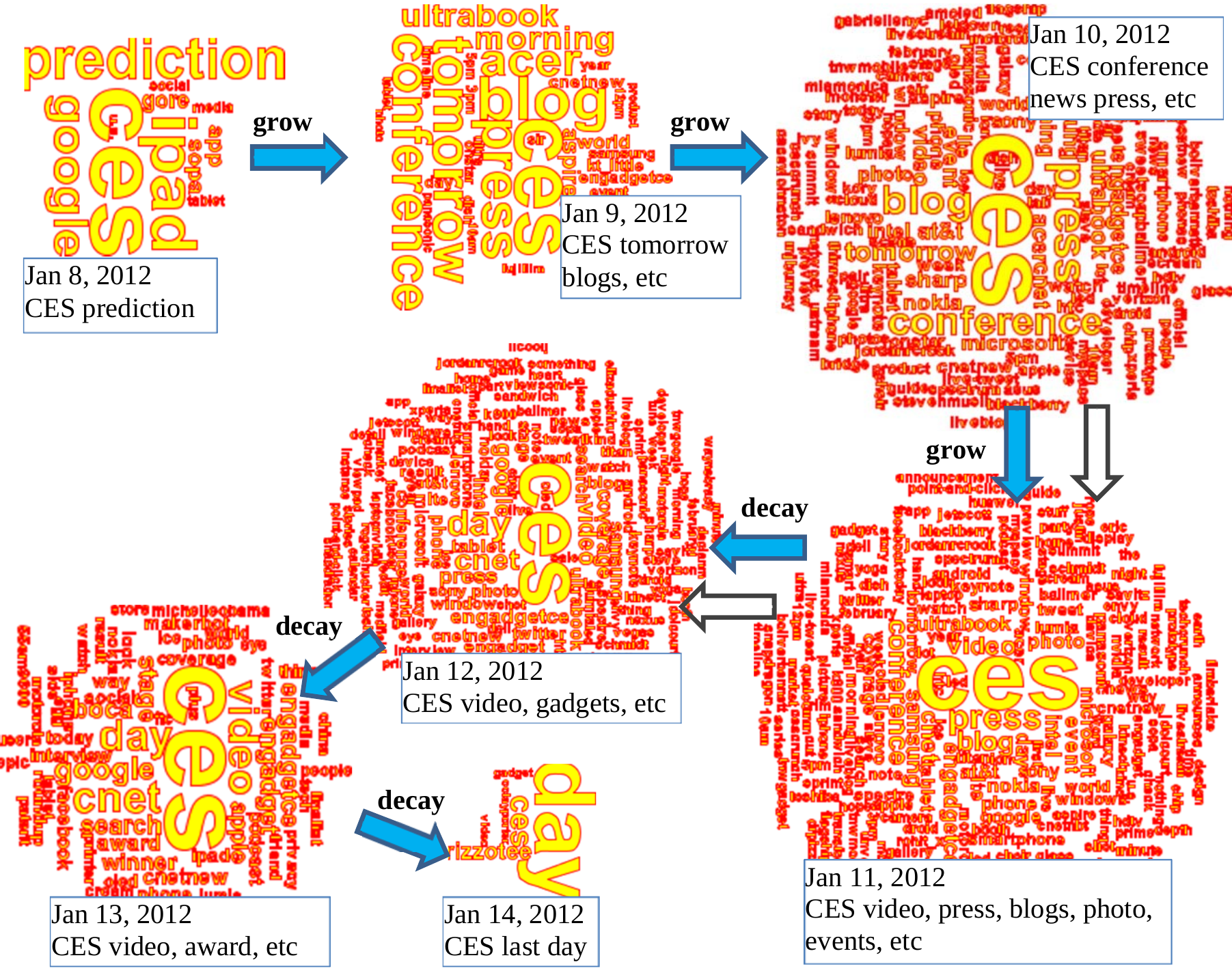}
  \label{evolution:a}
}\hspace{1cm}
\subfigure[The merging and splitting of ``SOPA'' and ``Apple'']{
  \includegraphics[height=6cm] {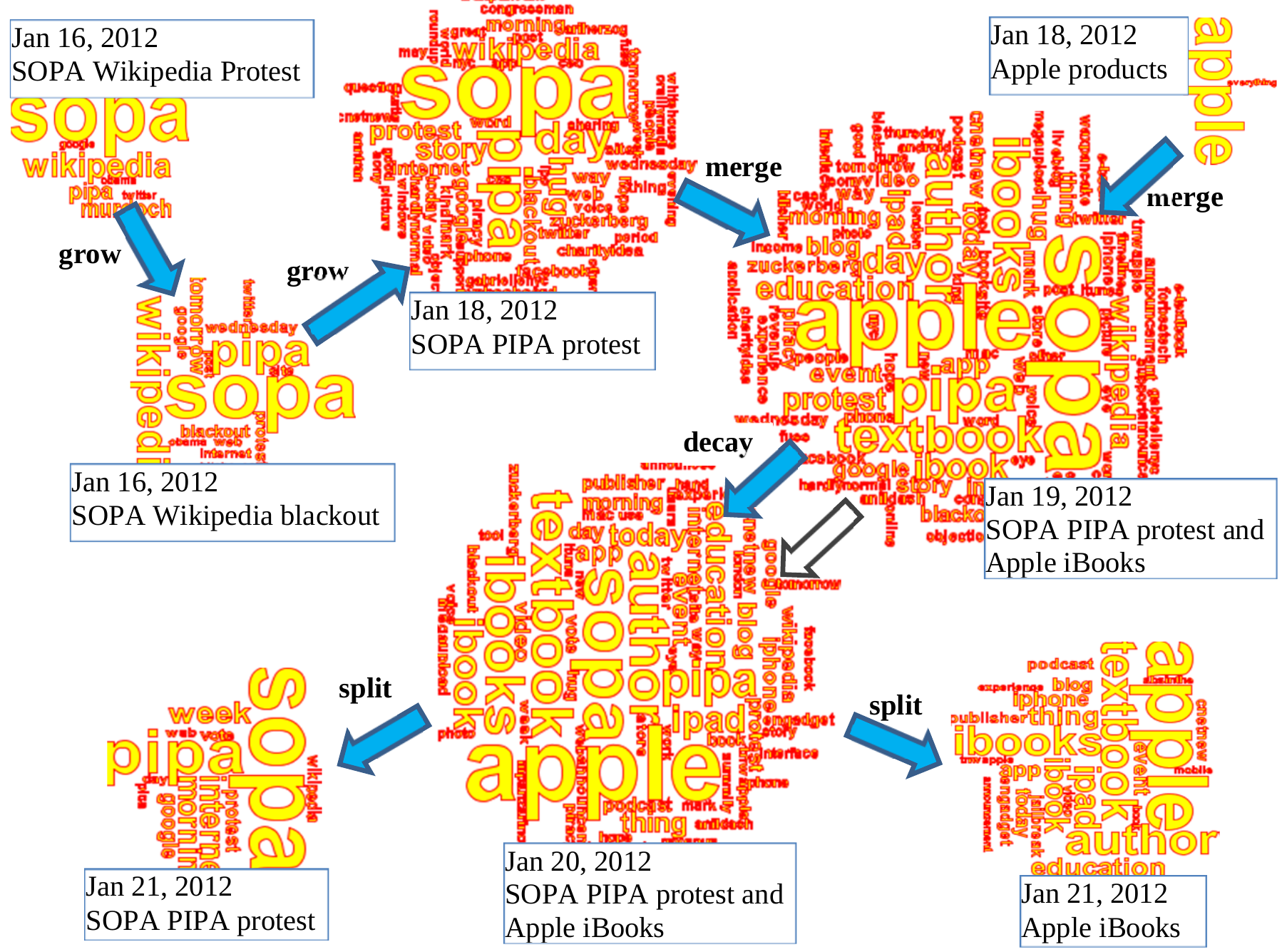}
  \label{evolution:b}
}\vspace{-0.2cm}
\caption{Tracking the evolution of events by \etrack and baselines.
At each moment, an event is annotated by a word cloud. Baselines 1 and 2 only works for the
detection of new emerging events, and is not applicable for the tracking of merging and splitting dynamics. The evolution trajectories of \etrack and Baseline 3 are depicted by solid and
hollow arrows respectively.
}\label{evolution}\vspace{-0.2cm}
\end{figure*}

Notice, both baselines above are for event detection only. Lists of the top 10 events
detected by \hashtags and \unigrams are presented in Figure \ref{top10} (first two columns).
Hashtags are employed by twitter users as a manual way to indicate an event, but it requires manual
assignation by a human. Some highly frequent hashtags like ``\#opengov'' and ``\#opendata'' are not
designed for event indication, hurting the precision. \unigrams uses the entities
extracted from the social stream preprocessing stage, which has a better quality than
\hashtags. However, both of them are limited in their representation of events, because the internal structure of events is missing. Besides, although these peaks can
detect bursty words, they cannot discover evolution patterns such as the merging/splitting 
of events. For example, in Figure \ref{googlepeaks}, there is no way to know ``Apple announced
iBooks'' is an event split from the big event ``SOPA'' earlier, as illustrated in detail in Figure
\ref{evolution:b}.

\stitle{Baseline 2: Community Detection}. A community in a large network refers
to a structure of nodes with dense connections internally and sparse connections between
communities. It is possible to define an event as a community of posts. Louvain method
\cite{louvain}, based on modularity optimization, is the state-of-the-art approach which outperforms
other known community detection methods in terms of performance. We design a baseline called
``\louvain'' to detect events defined as post communities. 

The top 10 events generated by
\louvain are shown in Figure \ref{top10}. As we can see, not every result is meaningful in
\louvain method. For example, ``Apple iphone ipad'' and ``Internet people
time'' are too vague to correspond to any concrete real events. The reason is, although \louvain
method can make sure every community has relatively dense internal and sparse external
connections, it cannot guarantee that every node in the community is important and has a
sufficiently high connectivity with other nodes in the same community. It is highly possible that a
low-degree node belongs to a community only because it has zero connectivity with other
communities. Furthermore, noise posts are quite prevalent in Twitter and they negatively
impact Louvain method.

\stitle{Baseline 3: Pattern-Matching}. We design a baseline to track the evolution
patterns of events between snapshots. In graph mining, the ``divide-and-conquer'' approach of
decomposing the evolving graph into a series of snapshot graphs at each moment is a traditional way
to tackle evolving graph related problems
\cite{kdd07-AsurPU,pvldb09-KimH}. As an example, Kim et al. \cite{pvldb09-KimH} first cluster 
individual snapshots into quasi-cliques and then map them in adjacent snapshots over time. Inspired 
by this approach, we design a
baseline for event evolution tracking, which detects events from each
snapshot of post network independently, and then characterizes the
evolution of these events at consecutive moments, by identifying
certain heuristic patterns:

{\small
$\bullet$ If $\frac{|E_t\cap E_{t+1}|}{|E_t\cup E_{t+1}|}\geq
\kappa\%$ and $|E_t|\leq |E_{t+1}|$, $E_{t+1}=\uparrow E_{t}$;

$\bullet$ If $\frac{|E_t\cap E_{t+1}|}{|E_t\cup E_{t+1}|}\geq
\kappa\%$ and $|E_t|>|E_{t+1}|$, $E_{t+1}=\downarrow E_{t}$.
}

\noindent where $E_{t}$ and $E_{t+1}$ are any two events detected at moment
$t$ and $t+1$ respectively, $\kappa\%$ is the minimal commonality to
say $E_t$ and $E_{t+1}$ are snapshots of the same event.
A higher $\kappa\%$ will result in a higher precision but a lower
recall of the evolution tracking. Empirically we found that
we needed to set $\kappa\%=90\%$
to guarantee the quality. It is worth noting that this baseline
generates the same events as \etrack, but with a non-incremental evolution
tracking approach.

\stitle{Precision and Recall}. To measure the quality of event detection, we use
\hashtags, \unigrams and \louvain as baselines to compare with our Algorithm \etrack. It is
worth noting that Baseline 3 is designed for the tracking of event evolution
dynamics between moments, so we omit it here. We
compare the precision and recall of top 20 events generated by baselines and \etrack and show
the results in Table 2(c). On the ground truth of 20 major events obtained from news articles
in technology websites, \hashtags and \unigrams have rather low precision and recall
scores, because of their poor ability in capturing event bursts. Notice that the precision and
recall may be not equal even the sizes of extracted events and ground truth are equal, because
multiple extracted events may correspond to the same event in ground truth. \etrack outperforms the
baselines in both precision and recall. Since there are many events discussed in the social media
but not very noticeable in news websites, we also validate the precision of the generated events
using Google Trends. As we can see, \hashtags and \unigrams don't gain too much in Google
Trends validation, since the words they generate are less informative and not very event-indicating.
\etrack gains a precision of 95\% in Google Trends, where the only failed result is ``Samsung galaxy
nexus'', whose volume is steadily high without obvious peaks in Google Trends. The reason may be
social stream is more dynamic. \louvain is worse than \etrack. The results
show \etrack is significantly better than other baselines in quality.

\stitle{Life Cycle of Event Evolution}. Our approach is capable of
tracking the whole life cycle of an event: from birth, growth, decay to death. We illustrate this
using the example of ``CES 2012'', a major consumer electronics
show held in Las Vegas from January 10 to 13. As early as Jan 6, our
approach has already detected some discussions about CES and generated
an event about CES. Figure \ref{evolution:a} shows the major snapshots of this event, from
Jan 8 to Jan 14. As we can see, on Jan 8, most
people talked about ``CES prediction'', and on Jan 9, the highlighted
topic was ``CES tomorrow'' and some hearsays about ``ultrabook''
which would be shown in CES. After the actual event happened on Jan
10, we can see the event grew distinctly bigger, and lots of products,
news and messages are spreading over the social network, and this
situation continues until Jan 13, which is the last day of CES.
Afterwards, the discussions become weaker and continue until
Jan 14, when ``CES'' was not the biggest mention on that day but still
existed in some discussions.
Compared with our approach, Baselines 1 and 2 can detect the emerging of ``CES'' with a frequency
count at each moment, but no trajectory is generated. Baseline 3 can track a very coarse trajectory of this
event, i.e., from Jan 10 to Jan 12. The reason is, if an event changes rapidly and many posts at
consecutive moments cannot be associated with each other, Baseline 3 will fail to track the
evolution. Since in social streams the posts usually surge quickly, our approach is 
superior over the baselines.

\stitle{Event Merging \& Splitting}.
Figure \ref{evolution:b} illustrates an example of event
merging and splitting generated by
Algorithm \etrack.
\etrack detected the event of SOPA (Stop Online Piracy Act) and Wikipedia on Jan 16, because on
that day Wikipedia announced the blackout on Wednesday (Jan 18) to
protest SOPA. This event grew distinctly on Jan 17 and Jan 18, by
inducing more people in the social network to discuss about this
topic. At the same time, there was another event detected on Jan 18,
discussing Apple's products. On Jan 19, actually the SOPA event and
Apple event were merged, because Apple joined the SOPA protest and
lots of Apple products such as iBooks in education are actually
directly related to SOPA. This event evolved on Jan 20, by adding more
discussions about iBooks 2. Apple iBooks 2 was actually unveiled in Jan
21, while this new product gained lots of attention, people who
talked about iBooks 
did not talk about SOPA anymore. Thus, on
Jan 21, the SOPA-Apple event was split into two events, which would
evolve independently afterwards.
Unfortunately, the above merging and
splitting process cannot be tracked by any of the baselines, which
output some independent events. The reason for Baseline 3 is, given the ground
truth $E_{t+1}=E_{t1}+E_{t2}$, i.e., $E_{t1}$ and $E_{t2}$ merged
into $E_{t+1}$,
$\frac{|E_{t1}\cap E_{t+1}|}{|E_{t1}\cup E_{t+1}|}\geq \kappa\%$ or
$\frac{|E_{t2}\cap E_{t+1}|}{|E_{t2}\cup E_{t+1}|}\geq \kappa\%$ is most likely invalid, so the
ground truth cannot be tracked.

\begin{figure}[t]
\centering
\subfigure[Varying time window]{
  \includegraphics[width=4cm] {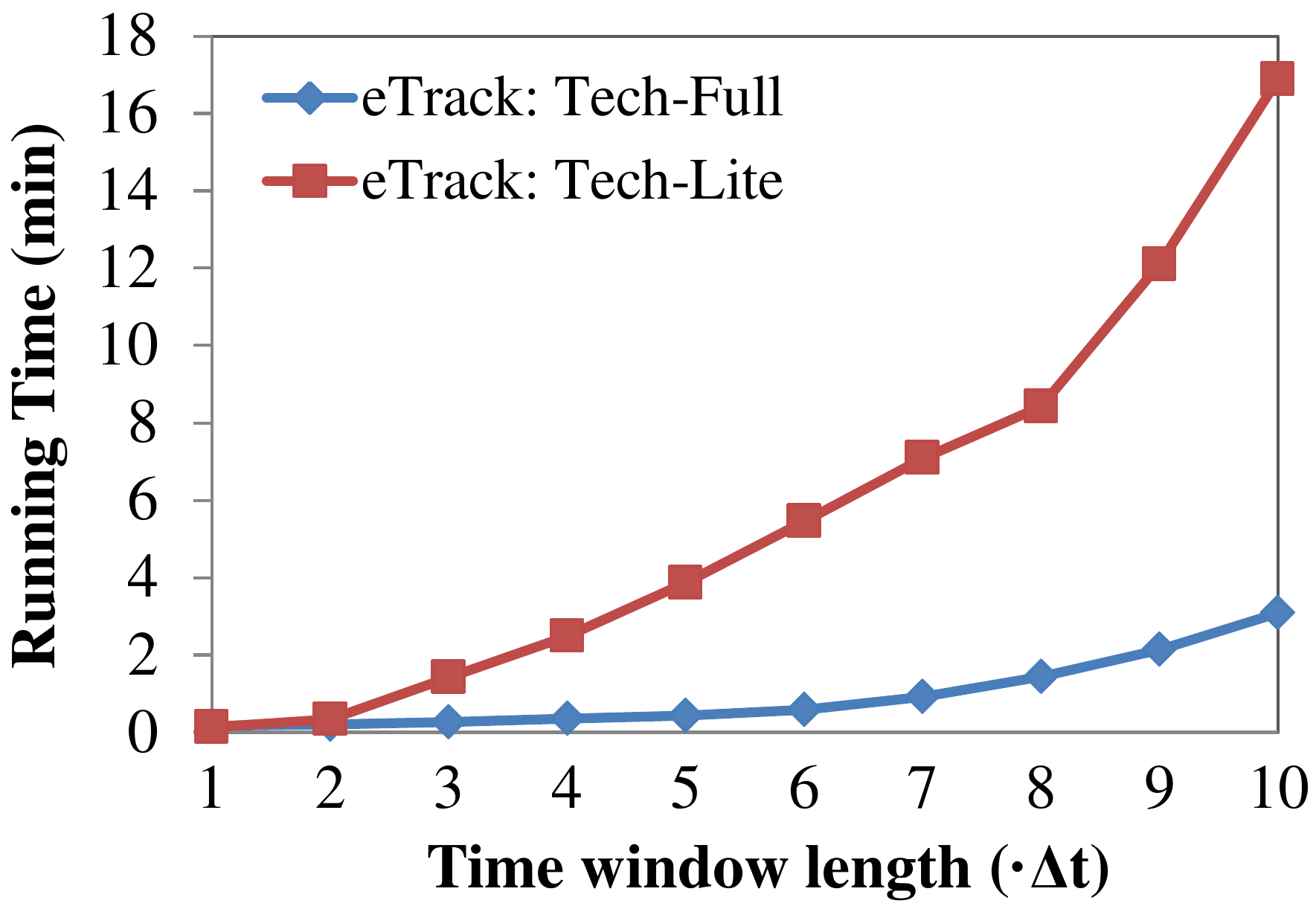}
  \label{performance:a}
}
\subfigure[Varying step length]{
  \includegraphics[width=4cm] {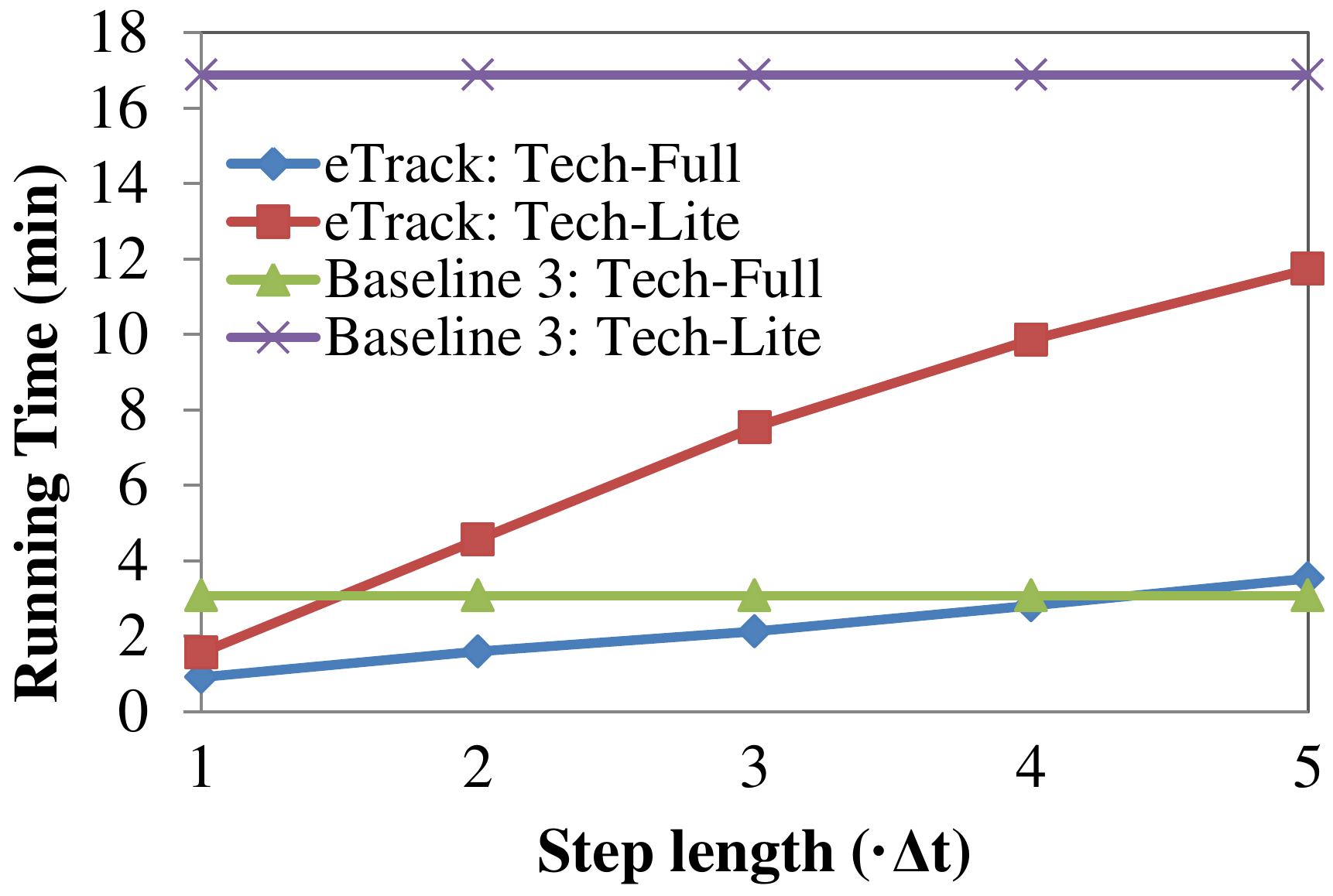}
  \label{performance:b}
}
\vspace{-0.3cm}\caption{The running time on two datasets as the adjusting of the time window length
and step length.}\label{performance}
\vspace{-0.2cm}\end{figure}
\subsection{Running Time}

We measure how the baseline and \etrack scale w.r.t. both the varying time
window width and the step length. 
We use both Tech-Lite
and Tech-Full streams, and set the time step interval $\Delta t=1$ day for
Tech-Lite, $\Delta t=1$ hour for Tech-Full to track events on different time granularity. The
streaming post rates for Tech-Lite and Tech-Full are 11700/day and 7126/hour respectively.
Figure \ref{performance:a}
shows the running time of \etrack when we increase the time window length, and we can see
for a time window of $10\Delta t$ hours in Tech-Full, our approach can finish the post
preprocessing, post network construction and event tracking in just 3 minutes. 
A key observation is that the running time of \etrack does not depend on the overall size of the
dataset. Rather, it depends on the streaming speed of posts in $\Delta t$.
Figure \ref{performance:b} shows if we fix the time window length as $10\Delta t$ and
increase the step length of the sliding time window, the running time of \etrack
grows nearly linearly. Compared with our
incremental computation, Baseline 3 has to process posts in the whole time window from scratch at
each moment, so the running time will be steadily high. 
If the step length is larger than $4\Delta t$ in TechFull, \etrack
does not have an advantage in running time compared with Baseline 3, because a large part of
post network is updated at each moment. However, this extreme case is rare. Since
in a real scenario, the step length is much smaller than the time window length, our approach shows much better efficiency than the baseline approach.